\documentclass[10pt,conference]{IEEEtran}

\usepackage[T1]{fontenc}
\usepackage[utf8]{inputenc}
\usepackage{cite}
\usepackage{amsmath,amssymb,amsfonts}
\usepackage{graphicx}
\usepackage{booktabs}
\usepackage{multirow}
\usepackage{array}
\usepackage{tabularx}
\usepackage{url}
\usepackage{balance}
\usepackage{xcolor}
\newif\ifshowrevisions
\showrevisionsfalse
\definecolor{revisioncolor}{RGB}{160,0,0}
\DeclareRobustCommand{\rev}[1]{%
  \ifshowrevisions\textcolor{revisioncolor}{#1}\else#1\fi
}
\usepackage{float}
\usepackage[most]{tcolorbox}
\usepackage{enumitem}
\usepackage{comment}
\usepackage{capt-of}
\usepackage{placeins}
\usepackage[colorlinks,urlcolor=blue,linkcolor=blue,citecolor=blue]{hyperref}
\graphicspath{{figures/}}

\newcommand{\tabsmall}{%
  \footnotesize
  \setlength{\tabcolsep}{3.5pt}%
}

\newcommand{\tabtiny}{%
  \scriptsize
  \setlength{\tabcolsep}{2.6pt}%
}

\newcommand{\tabnote}[1]{%
  \par\vspace{1.5pt}%
  \noindent\parbox{\linewidth}{\scriptsize #1}%
}

\newcommand{\dtime}{\Delta_{\mathrm{time}}}
\newcommand{\dmap}{\Delta_{\mathrm{map}}}
\newcommand{\dspread}{\Delta_{\mathrm{spread}}}

\newcommand{\pv}[1]{%
  \edef\pvtemp{#1}%
  \expandafter\pvparse\pvtemp\relax
}
\def\pvparse#1e#2\relax{%
  #1\times10^{\number\numexpr#2\relax}%
}

\newcommand{\FNScrVariants}{27}%                         % configurations in the screening sweep
\newcommand{\FNScrBlocks}{20}%                           % dataset x seed blocks
\newcommand{\FNScrRank}{4.03}%                           % LP/latency mean rank
\newcommand{\FNScrChi}{263.9}%                           % Friedman chi2
\newcommand{\FNScrP}{3.2e-41}%                           % Friedman p
\newcommand{\FNScrCD}{9.28}%                             % Nemenyi CD
\newcommand{\FNScrWithinCD}{14}%                         % configurations within one CD of the leader
\newcommand{\FNScrWinsA}{16}%                            % LP/latency wins vs LP/rate
\newcommand{\FNScrWinsB}{4}%                             % LP/rate wins
\newcommand{\FNScrMedDelta}{+0.70}%                      % median delta pp
\newcommand{\FNScrHolmP}{0.0084}%                        % Holm-adjusted p
\newcommand{\FNScrBootLo}{+0.11}%                        % bootstrap CI low
\newcommand{\FNScrBootHi}{+10.19}%                       % bootstrap CI high
\newcommand{\FNEncBlocks}{25}%                           % protocol x seed blocks
\newcommand{\FNEncLevels}{3}%                            % encodings compared
\newcommand{\FNEncRank}{1.44}%                           % champion mean rank
\newcommand{\FNEncP}{1.5e-03}%                           % Friedman p
\newcommand{\FNEncWins}{18}%                             % blocks won by the champion
\newcommand{\FNNeuBlocks}{20}%                           % protocol x seed blocks
\newcommand{\FNNeuLevels}{9}%                            % neurons compared
\newcommand{\FNNeuRank}{2.70}%                           % champion mean rank
\newcommand{\FNNeuP}{1.0e-06}%                           % Friedman p
\newcommand{\FNNeuRunnerUp}{Synaptic}%                   % second-ranked neuron
\newcommand{\FNNeuRunnerUpRank}{3.45}%                   % its mean rank
\newcommand{\FNNeuMostWins}{Leaky}%                      % neuron winning the most individual blocks
\newcommand{\FNTNNNslLat}{3}%                            % median T99, nslkdd_v2 latency
\newcommand{\FNTNFNslLat}{3}%                            % median T95, nslkdd_v2 latency
\newcommand{\FNTNNNslRat}{20}%                           % median T99, nslkdd_v2 rate
\newcommand{\FNTNFNslRat}{12}%                           % median T95, nslkdd_v2 rate
\newcommand{\FNTNNKddLat}{19}%                           % median T99, kddcup99_v2 latency
\newcommand{\FNTNFKddLat}{7}%                            % median T95, kddcup99_v2 latency
\newcommand{\FNTNNKddRat}{14}%                           % median T99, kddcup99_v2 rate
\newcommand{\FNTNFKddRat}{7}%                            % median T95, kddcup99_v2 rate
\newcommand{\FNTNNCicLat}{23}%                           % median T99, cicids2017_v2 latency
\newcommand{\FNTNFCicLat}{22}%                           % median T95, cicids2017_v2 latency
\newcommand{\FNTNNCicRat}{21}%                           % median T99, cicids2017_v2 rate
\newcommand{\FNTNFCicRat}{12}%                           % median T95, cicids2017_v2 rate
\newcommand{\FNTNNCtuALat}{14}%                          % median T99, ctu13_v2_f0 latency
\newcommand{\FNTNFCtuALat}{11}%                          % median T95, ctu13_v2_f0 latency
\newcommand{\FNTNNCtuARat}{17}%                          % median T99, ctu13_v2_f0 rate
\newcommand{\FNTNFCtuARat}{8}%                           % median T95, ctu13_v2_f0 rate
\newcommand{\FNTNNCtuBLat}{15}%                          % median T99, ctu13_v2_f1 latency
\newcommand{\FNTNFCtuBLat}{15}%                          % median T95, ctu13_v2_f1 latency
\newcommand{\FNTNNCtuBRat}{6}%                           % median T99, ctu13_v2_f1 rate
\newcommand{\FNTNFCtuBRat}{4}%                           % median T95, ctu13_v2_f1 rate
\newcommand{\FNSopRatioMin}{7.2}%                        % min rate/latency sops_mean
\newcommand{\FNSopRatioMax}{13.0}%                       % max rate/latency sops_mean
\newcommand{\FNInpRatioMin}{9.3}%                        % min rate/latency input_spikes
\newcommand{\FNInpRatioMax}{19.7}%                       % max rate/latency input_spikes
\newcommand{\FNTotRatioMin}{1.7}%                        % min rate/latency total_spikes
\newcommand{\FNTotRatioMax}{2.5}%                        % max rate/latency total_spikes
\newcommand{\FNCicFixedLat}{0.7842}%                     % fixed-universe macro-F1
\newcommand{\FNCicPresentLat}{0.9803}%                   % test-present macro-F1
\newcommand{\FNCicFixedRat}{0.7277}%                     % fixed-universe macro-F1
\newcommand{\FNCicFixedDel}{0.4981}%                     % fixed-universe macro-F1
\newcommand{\FNCicCeiling}{0.80}%                        % macro-F1 ceiling
\newcommand{\FNDupKdd}{57.53}%                           % percent KDDCup99 test rows duplicated in train
\newcommand{\FNDupNsl}{2.68}%                            % same, NSL-KDD
\newcommand{\FNDedupProtocols}{5}%                       % protocols with duplicates, encoding axis
\newcommand{\FNDedupKddOffMedian}{+1.045}%               % LP lat-vs-rate MEDIAN paired margin, official
\newcommand{\FNDedupKddDedMedian}{+0.976}%               % same, deduplicated
\newcommand{\FNFarValNsl}{1.0}%                          % realised validation FAR at a 1 percent target
\newcommand{\FNFarTestNsl}{7.7}%                         % realised test FAR
\newcommand{\FNFarValKdd}{1.0}%                          % realised validation FAR at a 1 percent target
\newcommand{\FNFarTestKdd}{1.9}%                         % realised test FAR
\newcommand{\FNFarValCic}{1.1}%                          % realised validation FAR at a 1 percent target
\newcommand{\FNFarTestCic}{1.2}%                         % realised test FAR
\newcommand{\FNFarValCtuA}{1.4}%                         % realised validation FAR at a 1 percent target
\newcommand{\FNFarTestCtuA}{3.6}%                        % realised test FAR
\newcommand{\FNFarValCtuB}{1.4}%                         % realised validation FAR at a 1 percent target
\newcommand{\FNFarTestCtuB}{48.0}%                       % realised test FAR
\newcommand{\FNKnownLo}{0.72}%                           % lowest known-type DR
\newcommand{\FNKnownHi}{0.80}%                           % highest known-type DR
\newcommand{\FNUnseenLo}{0.26}%                          % lowest unseen-type DR
\newcommand{\FNUnseenHi}{0.43}%                          % highest unseen-type DR
\newcommand{\FNUnseenFrac}{29.2}%                        % percent of KDDTest+ attack rows of unseen type
\newcommand{\FNUnseenConfigs}{27}%                       % configurations entering the known-vs-unseen comparison
\newcommand{\FNScrLeakyP}{0.33}%                         % LP/latency vs Leaky/latency, Wilcoxon p
\newcommand{\FNScrEncP}{0.027}%                          % encoding-level latency vs rate, Holm p
\newcommand{\FNScrEncCiLo}{-0.18}%                       % its CI low
\newcommand{\FNScrEncCiHi}{+11.30}%                      % its CI high
\newcommand{\FNKddAccRat}{0.9260}%                       % KDDCup99 accuracy, LP/rate
\newcommand{\FNKddAccLat}{0.9257}%                       % same, LP/latency
\newcommand{\FNKddFOneGain}{0.379}%                      % LP latency-minus-rate macro-F1 pp on KDDCup99
\newcommand{\FNDeltaDeficitNsl}{6.5}%                    % delta terminal deficit vs best on NSL-KDD, pp
\newcommand{\FNLatThresh}{0.01}%                         % latency encoder threshold
\newcommand{\FNDeltaThresh}{0.05}%                       % delta encoder threshold
\newcommand{\FNNslLatInSpk}{14.58}%                      % NSL-KDD latency input spikes per sample
\newcommand{\FNZeroNsl}{110.42}%                         % features exactly zero, nslkdd_v2
\newcommand{\FNBandNsl}{0.0000}%                         % features in (0, theta_lat], nslkdd_v2
\newcommand{\FNGateBandNsl}{1.020}%                      % features in (theta_lat, theta_delta], nslkdd_v2
\newcommand{\FNGateBandPctNsl}{6.99}%                    % that band as a percent of active features, nslkdd_v2
\newcommand{\FNDimNsl}{125}%                             % feature dimension, nslkdd_v2
\newcommand{\FNZeroKdd}{108.51}%                         % features exactly zero, kddcup99_v2
\newcommand{\FNBandKdd}{0.0000}%                         % features in (0, theta_lat], kddcup99_v2
\newcommand{\FNGateBandKdd}{0.404}%                      % features in (theta_lat, theta_delta], kddcup99_v2
\newcommand{\FNGateBandPctKdd}{3.24}%                    % that band as a percent of active features, kddcup99_v2
\newcommand{\FNDimKdd}{121}%                             % feature dimension, kddcup99_v2
\newcommand{\FNZeroCic}{38.03}%                          % features exactly zero, cicids2017_v2
\newcommand{\FNBandCic}{0.0637}%                         % features in (0, theta_lat], cicids2017_v2
\newcommand{\FNGateBandCic}{1.943}%                      % features in (theta_lat, theta_delta], cicids2017_v2
\newcommand{\FNGateBandPctCic}{4.86}%                    % that band as a percent of active features, cicids2017_v2
\newcommand{\FNDimCic}{78}%                              % feature dimension, cicids2017_v2
\newcommand{\FNZeroCtuA}{2.14}%                          % features exactly zero, ctu13_v2_f0
\newcommand{\FNBandCtuA}{0.6263}%                        % features in (0, theta_lat], ctu13_v2_f0
\newcommand{\FNGateBandCtuA}{0.870}%                     % features in (theta_lat, theta_delta], ctu13_v2_f0
\newcommand{\FNGateBandPctCtuA}{3.81}%                   % that band as a percent of active features, ctu13_v2_f0
\newcommand{\FNDimCtuA}{25}%                             % feature dimension, ctu13_v2_f0
\newcommand{\FNZeroCtuB}{4.00}%                          % features exactly zero, ctu13_v2_f1
\newcommand{\FNBandCtuB}{0.3475}%                        % features in (0, theta_lat], ctu13_v2_f1
\newcommand{\FNGateBandCtuB}{1.185}%                     % features in (theta_lat, theta_delta], ctu13_v2_f1
\newcommand{\FNGateBandPctCtuB}{5.64}%                   % that band as a percent of active features, ctu13_v2_f1
\newcommand{\FNDimCtuB}{25}%                             % feature dimension, ctu13_v2_f1
\newcommand{\FNNslTestN}{22{,}544}%                      % NSL-KDD confirmation test samples
\newcommand{\FNDimNslScreen}{122}%                       % NSL-KDD dimension under the screening protocol
\newcommand{\FNKddRateDeltaMarginScr}{0.21}%             % KDDCup99 rate-minus-delta mean margin, pp
\newcommand{\FNKddRateWinsScr}{2}%                       % seed-blocks where rate leads
\newcommand{\FNKddDeltaWinsScr}{3}%                      % seed-blocks where delta leads
\newcommand{\FNKddRateDeltaLoScr}{-3.1}%                 % most negative per-seed diff, pp
\newcommand{\FNKddRateDeltaHiScr}{+3.0}%                 % most positive per-seed diff, pp
\newcommand{\FNKddRateDeltaMarginConf}{+0.06}%           % KDDCup99 rate-minus-delta margin under the confirmation protocol, pp
\newcommand{\FNNslRtlRec}{0.164}%                        % NSL-KDD R2L recall
\newcommand{\FNNslUtrRec}{0.197}%                        % NSL-KDD U2R recall
\newcommand{\FNNslMacro}{0.5846}%                        % NSL-KDD macro-F1
\newcommand{\FNTimingNsl}{+1.17}%                         % timing premium at matched input spikes, nslkdd_v2
\newcommand{\FNTimingKdd}{-0.06}%                         % timing premium at matched input spikes, kddcup99_v2
\newcommand{\FNTimingCic}{+29.99}%                        % timing premium at matched input spikes, cicids2017_v2
\newcommand{\FNTimingCtuA}{+54.52}%                       % timing premium at matched input spikes, ctu13_v2_f0
\newcommand{\FNTimingCtuB}{-9.02}%                        % timing premium at matched input spikes, ctu13_v2_f1
\newcommand{\FNTimingSpread}{63.5}%                      % spread of the timing premium, pp
\newcommand{\FNConfBlocks}{25}%                          % confirmation protocol x seed blocks
\newcommand{\FNConfVariants}{27}%                        % configurations in the confirmation sweep
\newcommand{\FNConfChi}{137.0}%                          % Friedman chi2, confirmation
\newcommand{\FNConfP}{4.7e-17}%                          % Friedman p, confirmation
\newcommand{\FNConfCD}{8.30}%                            % Nemenyi CD, confirmation
\newcommand{\FNTau}{0.8063}%                             % Kendall tau, screening vs confirmation ranking
\newcommand{\FNTauP}{<0.0001}%                           % permutation p for that tau (bounded by 1/N)
\newcommand{\FNTauPerms}{10{,}000}%                      % permutations in the tau test
\newcommand{\FNArmLatNsl}{0.5818}%                       % A1 latency@0.01 macro-F1, 3 seeds, nslkdd_v2
\newcommand{\FNArmPermNsl}{0.5113}%                      % A2 permuted-time macro-F1, nslkdd_v2
\newcommand{\FNArmDeltaNsl}{0.5701}%                     % A3 delta@0.01 macro-F1, nslkdd_v2
\newcommand{\FNDmapNsl}{+7.04}%                          % value of the amplitude-to-time mapping, pp, nslkdd_v2
\newcommand{\FNDspreadNsl}{-5.87}%                       % value of temporal spreading, pp, nslkdd_v2
\newcommand{\FNDtimeNsl}{+1.17}%                         % timing premium at two decimals so the row sums, nslkdd_v2
\newcommand{\FNArmLatKdd}{0.6394}%                       % A1 latency@0.01 macro-F1, 3 seeds, kddcup99_v2
\newcommand{\FNArmPermKdd}{0.5824}%                      % A2 permuted-time macro-F1, kddcup99_v2
\newcommand{\FNArmDeltaKdd}{0.6400}%                     % A3 delta@0.01 macro-F1, kddcup99_v2
\newcommand{\FNDmapKdd}{+5.70}%                          % value of the amplitude-to-time mapping, pp, kddcup99_v2
\newcommand{\FNDspreadKdd}{-5.76}%                       % value of temporal spreading, pp, kddcup99_v2
\newcommand{\FNDtimeKdd}{-0.06}%                         % timing premium at two decimals so the row sums, kddcup99_v2
\newcommand{\FNArmLatCic}{0.7825}%                       % A1 latency@0.01 macro-F1, 3 seeds, cicids2017_v2
\newcommand{\FNArmPermCic}{0.6187}%                      % A2 permuted-time macro-F1, cicids2017_v2
\newcommand{\FNArmDeltaCic}{0.4826}%                     % A3 delta@0.01 macro-F1, cicids2017_v2
\newcommand{\FNDmapCic}{+16.38}%                         % value of the amplitude-to-time mapping, pp, cicids2017_v2
\newcommand{\FNDspreadCic}{+13.61}%                      % value of temporal spreading, pp, cicids2017_v2
\newcommand{\FNDtimeCic}{+29.99}%                        % timing premium at two decimals so the row sums, cicids2017_v2
\newcommand{\FNArmLatCtuA}{0.9995}%                      % A1 latency@0.01 macro-F1, 3 seeds, ctu13_v2_f0
\newcommand{\FNArmPermCtuA}{0.7603}%                     % A2 permuted-time macro-F1, ctu13_v2_f0
\newcommand{\FNArmDeltaCtuA}{0.4543}%                    % A3 delta@0.01 macro-F1, ctu13_v2_f0
\newcommand{\FNDmapCtuA}{+23.92}%                        % value of the amplitude-to-time mapping, pp, ctu13_v2_f0
\newcommand{\FNDspreadCtuA}{+30.60}%                     % value of temporal spreading, pp, ctu13_v2_f0
\newcommand{\FNDtimeCtuA}{+54.52}%                       % timing premium at two decimals so the row sums, ctu13_v2_f0
\newcommand{\FNArmLatCtuB}{0.7710}%                      % A1 latency@0.01 macro-F1, 3 seeds, ctu13_v2_f1
\newcommand{\FNArmPermCtuB}{0.7339}%                     % A2 permuted-time macro-F1, ctu13_v2_f1
\newcommand{\FNArmDeltaCtuB}{0.8612}%                    % A3 delta@0.01 macro-F1, ctu13_v2_f1
\newcommand{\FNDmapCtuB}{+3.71}%                         % value of the amplitude-to-time mapping, pp, ctu13_v2_f1
\newcommand{\FNDspreadCtuB}{-12.73}%                     % value of temporal spreading, pp, ctu13_v2_f1
\newcommand{\FNDtimeCtuB}{-9.02}%                        % timing premium at two decimals so the row sums, ctu13_v2_f1
\newcommand{\FNDmapMin}{+3.71}%                          % smallest mapping term
\newcommand{\FNDmapMax}{+23.92}%                         % largest mapping term
\newcommand{\FNDspreadMin}{-12.73}%                      % smallest spreading term
\newcommand{\FNDspreadMax}{+30.60}%                      % largest spreading term
\newcommand{\FNDeltaCicNormal}{0.9837}%                  % F1, cicids2017_v2 normal delta
\newcommand{\FNDeltaCicNormalSd}{0.0007}%                % its across-seed SD
\newcommand{\FNDeltaCicDos}{0.5167}%                     % F1, cicids2017_v2 dos delta
\newcommand{\FNDeltaCicDosSd}{0.0434}%                   % its across-seed SD
\newcommand{\FNDeltaCicProbe}{0.9900}%                   % F1, cicids2017_v2 probe delta
\newcommand{\FNDeltaCicProbeSd}{0.0000}%                 % its across-seed SD
\newcommand{\FNDeltaCicRtl}{0.0000}%                     % F1, cicids2017_v2 r2l delta
\newcommand{\FNDeltaCicRtlSd}{0.0000}%                   % its across-seed SD
\newcommand{\FNDeltaCicUtr}{0.0000}%                     % F1, cicids2017_v2 u2r delta
\newcommand{\FNLatCicNormal}{0.9987}%                    % F1, cicids2017_v2 normal latency
\newcommand{\FNLatCicNormalSd}{0.0002}%                  % its across-seed SD
\newcommand{\FNLatCicDos}{0.9649}%                       % F1, cicids2017_v2 dos latency
\newcommand{\FNLatCicDosSd}{0.0069}%                     % its across-seed SD
\newcommand{\FNLatCicProbe}{0.9971}%                     % F1, cicids2017_v2 probe latency
\newcommand{\FNLatCicProbeSd}{0.0002}%                   % its across-seed SD
\newcommand{\FNLatCicRtl}{0.9605}%                       % F1, cicids2017_v2 r2l latency
\newcommand{\FNLatCicRtlSd}{0.0123}%                     % its across-seed SD
\newcommand{\FNLatCicUtr}{0.0000}%                       % F1, cicids2017_v2 u2r latency
\newcommand{\FNRateCicNormal}{0.9958}%                   % F1, cicids2017_v2 normal rate
\newcommand{\FNRateCicNormalSd}{0.0004}%                 % its across-seed SD
\newcommand{\FNRateCicDos}{0.8989}%                      % F1, cicids2017_v2 dos rate
\newcommand{\FNRateCicDosSd}{0.0121}%                    % its across-seed SD
\newcommand{\FNRateCicProbe}{0.9958}%                    % F1, cicids2017_v2 probe rate
\newcommand{\FNRateCicProbeSd}{0.0003}%                  % its across-seed SD
\newcommand{\FNRateCicRtl}{0.7479}%                      % F1, cicids2017_v2 r2l rate
\newcommand{\FNRateCicRtlSd}{0.0790}%                    % its across-seed SD
\newcommand{\FNRateCicUtr}{0.0000}%                      % F1, cicids2017_v2 u2r rate
\newcommand{\FNDeltaCtuANormal}{0.6619}%                 % F1, ctu13_v2_f0 normal delta
\newcommand{\FNDeltaCtuANormalSd}{0.0022}%               % its across-seed SD
\newcommand{\FNDeltaCtuABotnet}{0.5258}%                 % F1, ctu13_v2_f0 botnet delta
\newcommand{\FNDeltaCtuABotnetSd}{0.0074}%               % its across-seed SD
\newcommand{\FNLatCtuANormal}{0.9983}%                   % F1, ctu13_v2_f0 normal latency
\newcommand{\FNLatCtuANormalSd}{0.0015}%                 % its across-seed SD
\newcommand{\FNLatCtuABotnet}{0.9988}%                   % F1, ctu13_v2_f0 botnet latency
\newcommand{\FNLatCtuABotnetSd}{0.0011}%                 % its across-seed SD
\newcommand{\FNRateCtuANormal}{0.9856}%                  % F1, ctu13_v2_f0 normal rate
\newcommand{\FNRateCtuANormalSd}{0.0020}%                % its across-seed SD
\newcommand{\FNRateCtuABotnet}{0.9895}%                  % F1, ctu13_v2_f0 botnet rate
\newcommand{\FNRateCtuABotnetSd}{0.0015}%                % its across-seed SD
\newcommand{\FNDeltaCtuBNormal}{0.8688}%                 % F1, ctu13_v2_f1 normal delta
\newcommand{\FNDeltaCtuBNormalSd}{0.0066}%               % its across-seed SD
\newcommand{\FNDeltaCtuBBotnet}{0.8963}%                 % F1, ctu13_v2_f1 botnet delta
\newcommand{\FNDeltaCtuBBotnetSd}{0.0068}%               % its across-seed SD
\newcommand{\FNLatCtuBNormal}{0.7028}%                   % F1, ctu13_v2_f1 normal latency
\newcommand{\FNLatCtuBNormalSd}{0.0063}%                 % its across-seed SD
\newcommand{\FNLatCtuBBotnet}{0.8377}%                   % F1, ctu13_v2_f1 botnet latency
\newcommand{\FNLatCtuBBotnetSd}{0.0019}%                 % its across-seed SD
\newcommand{\FNRateCtuBNormal}{0.7107}%                  % F1, ctu13_v2_f1 normal rate
\newcommand{\FNRateCtuBNormalSd}{0.0073}%                % its across-seed SD
\newcommand{\FNRateCtuBBotnet}{0.8394}%                  % F1, ctu13_v2_f1 botnet rate
\newcommand{\FNRateCtuBBotnetSd}{0.0027}%                % its across-seed SD
\newcommand{\FNDeltaKddNormal}{0.8423}%                  % F1, kddcup99_v2 normal delta
\newcommand{\FNDeltaKddNormalSd}{0.0011}%                % its across-seed SD
\newcommand{\FNDeltaKddDos}{0.9831}%                     % F1, kddcup99_v2 dos delta
\newcommand{\FNDeltaKddDosSd}{0.0006}%                   % its across-seed SD
\newcommand{\FNDeltaKddProbe}{0.7501}%                   % F1, kddcup99_v2 probe delta
\newcommand{\FNDeltaKddProbeSd}{0.0211}%                 % its across-seed SD
\newcommand{\FNDeltaKddRtl}{0.0534}%                     % F1, kddcup99_v2 r2l delta
\newcommand{\FNDeltaKddRtlSd}{0.0188}%                   % its across-seed SD
\newcommand{\FNDeltaKddUtr}{0.4126}%                     % F1, kddcup99_v2 u2r delta
\newcommand{\FNDeltaKddUtrSd}{0.1828}%                   % its across-seed SD
\newcommand{\FNLatKddNormal}{0.8400}%                    % F1, kddcup99_v2 normal latency
\newcommand{\FNLatKddNormalSd}{0.0030}%                  % its across-seed SD
\newcommand{\FNLatKddDos}{0.9856}%                       % F1, kddcup99_v2 dos latency
\newcommand{\FNLatKddDosSd}{0.0005}%                     % its across-seed SD
\newcommand{\FNLatKddProbe}{0.7759}%                     % F1, kddcup99_v2 probe latency
\newcommand{\FNLatKddProbeSd}{0.0334}%                   % its across-seed SD
\newcommand{\FNLatKddRtl}{0.1136}%                       % F1, kddcup99_v2 r2l latency
\newcommand{\FNLatKddRtlSd}{0.0100}%                     % its across-seed SD
\newcommand{\FNLatKddUtr}{0.4397}%                       % F1, kddcup99_v2 u2r latency
\newcommand{\FNLatKddUtrSd}{0.0650}%                     % its across-seed SD
\newcommand{\FNRateKddNormal}{0.8414}%                   % F1, kddcup99_v2 normal rate
\newcommand{\FNRateKddNormalSd}{0.0035}%                 % its across-seed SD
\newcommand{\FNRateKddDos}{0.9856}%                      % F1, kddcup99_v2 dos rate
\newcommand{\FNRateKddDosSd}{0.0005}%                    % its across-seed SD
\newcommand{\FNRateKddProbe}{0.7831}%                    % F1, kddcup99_v2 probe rate
\newcommand{\FNRateKddProbeSd}{0.0324}%                  % its across-seed SD
\newcommand{\FNRateKddRtl}{0.0999}%                      % F1, kddcup99_v2 r2l rate
\newcommand{\FNRateKddRtlSd}{0.0167}%                    % its across-seed SD
\newcommand{\FNRateKddUtr}{0.3344}%                      % F1, kddcup99_v2 u2r rate
\newcommand{\FNRateKddUtrSd}{0.1627}%                    % its across-seed SD
\newcommand{\FNDeltaNslNormal}{0.7978}%                  % F1, nslkdd_v2 normal delta
\newcommand{\FNDeltaNslNormalSd}{0.0089}%                % its across-seed SD
\newcommand{\FNDeltaNslDos}{0.8514}%                     % F1, nslkdd_v2 dos delta
\newcommand{\FNDeltaNslDosSd}{0.0190}%                   % its across-seed SD
\newcommand{\FNDeltaNslProbe}{0.6878}%                   % F1, nslkdd_v2 probe delta
\newcommand{\FNDeltaNslProbeSd}{0.0275}%                 % its across-seed SD
\newcommand{\FNDeltaNslRtl}{0.0932}%                     % F1, nslkdd_v2 r2l delta
\newcommand{\FNDeltaNslRtlSd}{0.0167}%                   % its across-seed SD
\newcommand{\FNDeltaNslUtr}{0.1684}%                     % F1, nslkdd_v2 u2r delta
\newcommand{\FNDeltaNslUtrSd}{0.0788}%                   % its across-seed SD
\newcommand{\FNLatNslNormal}{0.7887}%                    % F1, nslkdd_v2 normal latency
\newcommand{\FNLatNslNormalSd}{0.0175}%                  % its across-seed SD
\newcommand{\FNLatNslDos}{0.8820}%                       % F1, nslkdd_v2 dos latency
\newcommand{\FNLatNslDosSd}{0.0157}%                     % its across-seed SD
\newcommand{\FNLatNslProbe}{0.6887}%                     % F1, nslkdd_v2 probe latency
\newcommand{\FNLatNslProbeSd}{0.0262}%                   % its across-seed SD
\newcommand{\FNLatNslRtl}{0.2753}%                       % F1, nslkdd_v2 r2l latency
\newcommand{\FNLatNslRtlSd}{0.0658}%                     % its across-seed SD
\newcommand{\FNLatNslUtr}{0.2883}%                       % F1, nslkdd_v2 u2r latency
\newcommand{\FNLatNslUtrSd}{0.1284}%                     % its across-seed SD
\newcommand{\FNRateNslNormal}{0.7877}%                   % F1, nslkdd_v2 normal rate
\newcommand{\FNRateNslNormalSd}{0.0077}%                 % its across-seed SD
\newcommand{\FNRateNslDos}{0.8836}%                      % F1, nslkdd_v2 dos rate
\newcommand{\FNRateNslDosSd}{0.0171}%                    % its across-seed SD
\newcommand{\FNRateNslProbe}{0.6909}%                    % F1, nslkdd_v2 probe rate
\newcommand{\FNRateNslProbeSd}{0.0259}%                  % its across-seed SD
\newcommand{\FNRateNslRtl}{0.1631}%                      % F1, nslkdd_v2 r2l rate
\newcommand{\FNRateNslRtlSd}{0.0306}%                    % its across-seed SD
\newcommand{\FNRateNslUtr}{0.2101}%                      % F1, nslkdd_v2 u2r rate
\newcommand{\FNRateNslUtrSd}{0.1732}%                    % its across-seed SD
\newcommand{\FNSupCicDos}{18{,}892}%                     % test support, cicids2017_v2 dos
\newcommand{\FNSupCicNormal}{523{,}965}%                 % test support, cicids2017_v2 normal
\newcommand{\FNSupCicProbe}{21{,}734}%                   % test support, cicids2017_v2 probe
\newcommand{\FNSupCicRtl}{1{,}561}%                      % test support, cicids2017_v2 r2l
\newcommand{\FNSupCicUtr}{0}%                            % test support, cicids2017_v2 u2r
\newcommand{\FNSupCtuABotnet}{212{,}710}%                % test support, ctu13_v2_f0 botnet
\newcommand{\FNSupCtuANormal}{151{,}533}%                % test support, ctu13_v2_f0 normal
\newcommand{\FNSupCtuBBotnet}{114{,}647}%                % test support, ctu13_v2_f1 botnet
\newcommand{\FNSupCtuBNormal}{96{,}297}%                 % test support, ctu13_v2_f1 normal
\newcommand{\FNSupKddDos}{229{,}855}%                    % test support, kddcup99_v2 dos
\newcommand{\FNSupKddNormal}{60{,}593}%                  % test support, kddcup99_v2 normal
\newcommand{\FNSupKddProbe}{4{,}166}%                    % test support, kddcup99_v2 probe
\newcommand{\FNSupKddRtl}{16{,}345}%                     % test support, kddcup99_v2 r2l
\newcommand{\FNSupKddUtr}{70}%                           % test support, kddcup99_v2 u2r
\newcommand{\FNSupNslDos}{7{,}460}%                      % test support, nslkdd_v2 dos
\newcommand{\FNSupNslNormal}{9{,}711}%                   % test support, nslkdd_v2 normal
\newcommand{\FNSupNslProbe}{2{,}421}%                    % test support, nslkdd_v2 probe
\newcommand{\FNSupNslRtl}{2{,}885}%                      % test support, nslkdd_v2 r2l
\newcommand{\FNSupNslUtr}{67}%                           % test support, nslkdd_v2 u2r
\newcommand{\FNTsweepCells}{20}%                         % protocol x T cells in the A2 sweep
\newcommand{\FNMapMovesN}{4}%                            % protocols where Delta_map moves with T beyond its own SE
\newcommand{\FNSpreadMovesN}{2}%                         % protocols where Delta_spread moves with T
\newcommand{\FNMapMeanNsl}{+5.29}%                       % mean Delta_map over T, nslkdd_v2
\newcommand{\FNSpreadMeanNsl}{-3.58}%                    % mean Delta_spread over T, nslkdd_v2
\newcommand{\FNDtimeMeanNsl}{+1.70}%                     % mean Delta_time over T, nslkdd_v2
\newcommand{\FNRhoNsl}{+0.06}%                           % corr(Delta_map, Delta_spread) across T, nslkdd_v2
\newcommand{\FNMapMeanKdd}{+3.42}%                       % mean Delta_map over T, kddcup99_v2
\newcommand{\FNSpreadMeanKdd}{-3.54}%                    % mean Delta_spread over T, kddcup99_v2
\newcommand{\FNDtimeMeanKdd}{-0.12}%                     % mean Delta_time over T, kddcup99_v2
\newcommand{\FNRhoKdd}{-0.93}%                           % corr(Delta_map, Delta_spread) across T, kddcup99_v2
\newcommand{\FNMapMeanCic}{+17.63}%                      % mean Delta_map over T, cicids2017_v2
\newcommand{\FNSpreadMeanCic}{+12.77}%                   % mean Delta_spread over T, cicids2017_v2
\newcommand{\FNDtimeMeanCic}{+30.39}%                    % mean Delta_time over T, cicids2017_v2
\newcommand{\FNRhoCic}{-0.91}%                           % corr(Delta_map, Delta_spread) across T, cicids2017_v2
\newcommand{\FNMapMeanCtuA}{+25.52}%                     % mean Delta_map over T, ctu13_v2_f0
\newcommand{\FNSpreadMeanCtuA}{+28.67}%                  % mean Delta_spread over T, ctu13_v2_f0
\newcommand{\FNDtimeMeanCtuA}{+54.19}%                   % mean Delta_time over T, ctu13_v2_f0
\newcommand{\FNRhoCtuA}{-0.94}%                          % corr(Delta_map, Delta_spread) across T, ctu13_v2_f0
\newcommand{\FNMapMeanCtuB}{+22.92}%                     % mean Delta_map over T, ctu13_v2_f1
\newcommand{\FNSpreadMeanCtuB}{-26.13}%                  % mean Delta_spread over T, ctu13_v2_f1
\newcommand{\FNDtimeMeanCtuB}{-3.21}%                    % mean Delta_time over T, ctu13_v2_f1
\newcommand{\FNRhoCtuB}{-0.92}%                          % corr(Delta_map, Delta_spread) across T, ctu13_v2_f1
\newcommand{\FNCtuBAtwoTfive}{0.2629}%                   % A2 macro-F1 on CTU-13 f1 at T=5
\newcommand{\FNCtuBAtwoTtwentyfive}{0.7339}%             % A2 macro-F1 on CTU-13 f1 at T=25
\newcommand{\FNCtuBIdenticalFrac}{0.31}%                 % share of active features keeping their own time under permutation
\newcommand{\FNCtuBSpearmanAtwo}{-0.026}%                % amplitude-time rank correlation of the permuted arm there
\newcommand{\FNKddIdenticalFrac}{0.70}%                  % same share on KDDCup99, the protocol where ties really do bind
\newcommand{\FNEoneSeeds}{three}%                        % seeds per cell in the matched-input arms
\newcommand{\FNSopResidual}{10^{-6}}%                    % tolerance of the SOP identity check
\newcommand{\FNCtuCausalCost}{0.31}%                     % pp lost to causal aggregation alone
\newcommand{\FNCtuScenarioCost}{7.68}%                   % pp lost to the scenario-disjoint split
\newcommand{\FNCtuTotalCost}{7.99}%                      % pp from the submitted protocol to the strict one
\newcommand{\FNConfRank}{5.92}%                          % mean rank of the selected configuration, confirmation
\newcommand{\FNConfLatTop}{four}%                        % latency configurations in the leading four
\newcommand{\FNConfEncLat}{11.44}%                       % encoding aggregation, latency
\newcommand{\FNConfEncRat}{13.75}%                       % same, rate
\newcommand{\FNConfEncDel}{16.80}%                       % same, delta
\newcommand{\FNConfEncWinsLat}{22}%                      % blocks won by latency
\newcommand{\FNConfEncWinsRat}{2}%                       % by rate
\newcommand{\FNConfEncWinsDel}{1}%                       % by delta
\newcommand{\FNATwoPredPass}{43.9}%                      % share of the CTU-13 f0 premium carried by the mapping, percent
\newcommand{\FNDmapRange}{20.20}%                        % spread of Delta_map across protocols at T=25, pp
\newcommand{\FNDspreadRange}{43.33}%                     % same for Delta_spread
\newcommand{\FNTsweepR}{+0.051}%                         % pooled Pearson(T, Delta_time) over the sweep
\newcommand{\FNTauTop}{0.7143}%                          % tau over the configurations within one screening CD
\newcommand{\FNTauBottom}{0.6923}%                       % tau over the remainder
\newcommand{\FNTauTopN}{14}%                             % configurations within one CD
\newcommand{\FNTauBottomN}{13}%                          % configurations outside one CD
\newcommand{\FNRBOninety}{0.8327}%                       % rank-biased overlap, p=0.9
\newcommand{\FNRBOeighty}{0.8870}%                       % rank-biased overlap, p=0.8
\newcommand{\FNTauAPscr}{0.7782}%                        % AP rank correlation, screening as reference
\newcommand{\FNTauAPconf}{0.7929}%                       % same, confirmation as reference
\newcommand{\FNTopKthree}{2}%                            % top-3 membership overlap
\newcommand{\FNTopKfive}{4}%                             % top-5 membership overlap
\newcommand{\FNTopKten}{9}%                              % top-10 membership overlap
\newcommand{\FNTopKfourteen}{12}%                        % top-14 membership overlap
\newcommand{\FNTauTopP}{1.6\times10^{-4}}%               % permutation p, leading group only
\newcommand{\FNTauTopPerms}{100{,}000}%                  % permutations in the leading-group test
\newcommand{\FNSpearmanLatency}{-0.99}%                  % median amplitude-time rank correlation, mapped arm
\newcommand{\FNSpearmanSpread}{-0.05}%                   % same, spread-only arm
\newcommand{\FNSpearmanLatencyLo}{-0.86}%                % weakest per-protocol value, mapped arm (KDDCup99)
\newcommand{\FNSpearmanSpreadLo}{-0.21}%                 % strongest residual correlation, spread-only arm (KDDCup99)
\newcommand{\FNParetoCostSpan}{58}%                      % rate/latency SOP span across all 27 configurations, NSL-KDD
\newcommand{\FNShiftKsCtuB}{0.4864}%                     % median per-feature KS, CTU-13 f1 (most shifted condition)
\newcommand{\FNShiftAucCtuB}{0.9787}%                    % train-vs-test domain-classifier AUC, CTU-13 f1
\newcommand{\FNShiftAucKddLo}{0.72}%                     % lowest domain AUC in the KDD family
\newcommand{\FNShiftAucKddHi}{0.75}%                     % highest domain AUC in the KDD family
\newcommand{\FNShiftRho}{0.000}%                         % Spearman(shift, timing premium) over the confirmation protocols
\newcommand{\FNNoCtuBlocks}{15}%                         % screening blocks excluding CTU-13
\newcommand{\FNNoCtuLeader}{Leaky/latency}%              % leader without CTU-13
\newcommand{\FNNoCtuLeadRank}{4.93}%                     % its mean rank
\newcommand{\FNNoCtuChampRank}{5.00}%                    % LP/latency mean rank without CTU-13
\newcommand{\FNNoCtuMargin}{0.067}%                      % margin between the two
\newcommand{\FNNoCtuLatTop}{five}%                       % latency configs in the leading five
\newcommand{\FNCtuIdxFA}{99.85}%                         % CTU-13 fold 0 mean, fold-index order
\newcommand{\FNCtuIdxFB}{77.02}%                         % CTU-13 fold 1 mean, fold-index order
\newcommand{\FNCtuIdxFC}{91.22}%                         % CTU-13 fold 2 mean, fold-index order
\newcommand{\FNCtuIdxFD}{99.95}%                         % CTU-13 fold 3 mean, fold-index order
\newcommand{\FNCtuNEval}{20}%                            % fold x seed evaluations behind the CTU-13 estimate
\newcommand{\FNCtuSdEval}{9.72}%                         % SD over those evaluations (ddof=1); NOT an across-fold SD
\newcommand{\FNFarDrNsl}{72.2}%                          % DR at the frozen 1% threshold, nslkdd_v2
\newcommand{\FNFarDrKdd}{91.9}%                          % DR at the frozen 1% threshold, kddcup99_v2
\newcommand{\FNFarDrCic}{99.5}%                          % DR at the frozen 1% threshold, cicids2017_v2
\newcommand{\FNFarDrCtuA}{100.0}%                        % DR at the frozen 1% threshold, ctu13_v2_f0
\newcommand{\FNFarDrCtuB}{100.0}%                        % DR at the frozen 1% threshold, ctu13_v2_f1
\newcommand{\FNCtuBNeuRSyn}{0.9114}%                     % RSynaptic macro-F1 on CTU-13 fold 1, neuron axis
\newcommand{\FNCtuBNeuRLky}{0.8984}%                     % RLeaky macro-F1 on CTU-13 fold 1, neuron axis
\newcommand{\FNCtuBNeuLP}{0.7702}%                       % LeakyParallel macro-F1 on CTU-13 fold 1, neuron axis
\newcommand{\FNCtuBEncDel}{0.8825}%                      % delta macro-F1 on CTU-13 fold 1, encoding axis
\newcommand{\FNCtuBEncRat}{0.7750}%                      % rate macro-F1 on CTU-13 fold 1, encoding axis
\newcommand{\FNCtuBEncLat}{0.7702}%                      % latency macro-F1 on CTU-13 fold 1, encoding axis
\newcommand{\FNCtuStrict}{92.01}%                        % four-fold strict CTU-13 estimate
\newcommand{\FNCtuLegacy}{100.00}%                       % submitted-protocol CTU-13 value
\newcommand{\FNCtuCausalRandom}{99.69}%                  % causal aggregates, random split
\newcommand{\FNSeeds}{5}%                                % seeds per cell

\newcommand{\FNNslMacroRate}{0.5471}
\newcommand{\FNNslMacroDelta}{0.5197}

\newcommand{\FNKddTestN}{311{,}029}
\newcommand{\FNKddMacro}{0.6309}
\newcommand{\FNKddMacroRate}{0.6089}
\newcommand{\FNKddMacroDelta}{0.6083}

\newcommand{\FNCicTestN}{566{,}152}

\newcommand{\FNCtuATestN}{364{,}243}
\newcommand{\FNCtuAEncLat}{0.9985}
\newcommand{\FNCtuAEncRat}{0.9876}
\newcommand{\FNCtuAEncDel}{0.5938}

\newcommand{\FNCtuBTestN}{210{,}944}

\newcommand{\FNDimCtuC}{25}
\newcommand{\FNDimCtuD}{25}

\tcbset{
  paperbox/.style={
    enhanced,
    breakable,
    colframe=gray!90!black,
    colback=black!5!white,
    boxrule=0.5pt,
    arc=3pt,
    left=0mm,
    right=1.5mm,
    bottom=1mm,
    top=3.5mm,
    before skip=6pt plus 2pt,
    after skip=6pt plus 2pt,
    fonttitle=\bfseries\small,
    coltitle=white,
    colbacktitle=gray!90!black,
    attach boxed title to top left={
      yshift=-3mm,
      xshift=2mm
    },
    boxed title style={
      boxrule=0.5pt,
      arc=1pt,
      left=1mm,
      right=1mm,
      top=0.4mm,
      bottom=0.4mm
    },
  }
}

\newtcolorbox{boxEnvRQ}{
  paperbox,
  title={Research Questions (RQs)}
}

\newtcolorbox{boxEnvContrib}{
  paperbox,
  title={Contributions}
}

\newlist{boxitems}{itemize}{1}
\setlist[boxitems]{
  label=\textbullet,
  nosep,
  leftmargin=0.9em,
  labelsep=0.35em,
  labelindent=0pt,
  itemindent=0pt,
  itemsep=1pt,
  topsep=0pt,
  parsep=0pt,
  partopsep=0pt
}

\begin{document}

\title{The Value of Spike Timing: \\ A Leakage-Resistant Benchmark of SNN Design Choices for Network Intrusion Detection}

% CR72: Add Sudip Mittal; preserve the relative order of the existing authors.
% AUTHOR CHECK: Sudip is placed before Shahram Rahimi, preserving the original
% final-author position. Confirm this insertion against the updated CITATION.cff.
% Affiliation/email: https://eng.ua.edu/eng-directory/dr-sudip-mittal/
% Numeric email keys avoid a collision between IEEE's seventh symbol (**)
% and the corresponding-author mark already used for Sayanton Dibbo.
\author{\IEEEauthorblockN{Raj Patel\textsuperscript{1},
Shaswata Mitra\textsuperscript{2},
David Amebley\textsuperscript{3},
Taye Akinrele\textsuperscript{4},\\
Sayanton Dibbo\textsuperscript{5,**},
\rev{Sudip Mittal\textsuperscript{6}},
Shahram Rahimi\textsuperscript{7}
}
 
\IEEEauthorblockA{Department of Computer Science,
The University of Alabama\\
\{rpatel38\textsuperscript{1},
smitra3\textsuperscript{2},
dkamebley\textsuperscript{3},
toakinrele\textsuperscript{4},
sdibbo\textsuperscript{5},
\rev{sudip.mittal\textsuperscript{6}},
srahimi1\textsuperscript{7}\}@ua.edu
}
}

\maketitle
\begingroup
\let\thefootnote\relax
\footnotetext{\textsuperscript{**}Corresponding author:
Sayanton Dibbo (sdibbo@ua.edu)}
\footnotetext{Code, run records, and analysis outputs
are available at
\url{https://github.com/rajpatel-UA/IntraSNNEval}.}
\endgroup
\begin{abstract}
    Spiking neural networks (SNNs) are increasingly studied for network intrusion
    detection, but comparative evidence on how neuron models and spike encodings
    affect performance remains limited. Evaluation choices can further influence
    the observed results when preprocessing, capture structure, or scenario
    information crosses the train--test boundary. We evaluate nine
    \texttt{snnTorch} neuron families with three spike encodings, yielding 27 SNN
    configurations across four intrusion-detection benchmarks. We first screen the
    complete design space and then repeat the evaluation using train-only
    transforms, seed-independent partitions, and capture- or scenario-aware
% CR01: Qualify the abstract ranking claim
    separation. \rev{At the shared operating point,
    \texttt{LeakyParallel/latency} achieved the lowest mean rank in both sweeps
    but was not statistically separated from \texttt{Leaky/latency}; their
    screening order reversed when CTU-13 was excluded. Overall, screening and
    confirmation retained substantial rank agreement despite changes in measured
    performance.}
    
    We then examine what latency coding contributes by holding the active spike set
    fixed and changing only its temporal organization. At the main $T=25$
    operating point, mapping feature magnitude to spike time improved macro-F1 on
    all five confirmation protocols, while spreading the same spikes through time
    without that mapping could either improve or reduce performance. 
% CR40: Tighten the abstract mechanism summary
\rev{Latency coding thus combines informative spike timing with temporal network
    dynamics. Sparse input encoding does not directly imply sparse internal
    activity, although latency coding requires fewer fanout-weighted operations
    than rate coding in the evaluated models. These findings motivate separate
    assessment of SNN design, evaluation protocol, spike timing, and computational
    activity for static network-flow data.}
\end{abstract}


\section{Introduction}\label{sec:introduction}
Network intrusion detection systems (NIDS) increasingly use learned models to
distinguish benign from malicious network traffic. Spiking neural networks
(SNNs) offer an alternative computational model in which information is
represented through discrete spike events and temporal state, making them
relevant to event-driven and neuromorphic computing
~\cite{roy2019towards,pfeiffer2018deep,rathi2023neuromorphic,davies2018loihi}.
For network-flow classification, however, the input is typically a static
feature vector rather than an event stream. The temporal structure presented to
an SNN must therefore be introduced by the encoding process. Consequently, the
choice of spike encoding becomes part of the representation itself and must be
considered together with the neuron dynamics that process it.

Prior SNN studies in network intrusion and anomaly detection have examined
temporal coding, convolutional architectures, and applications across
conventional network, automotive, and industrial-control traffic
~\cite{zhou2021spiking,wang2024efficient,knapp2025efficacy,
jaoudi2020carhacking,demertzis2020gryphon}. These studies establish the feasibility of
SNN-based NIDS, but most evaluate a selected architecture and encoding rather
than comparing multiple neuron and encoding choices within a common training
pipeline. This makes it difficult to determine how much of an observed result
is associated with the neuron model, the spike representation, or the
evaluation procedure. The latter is especially important for established NIDS
benchmarks, where exact duplicates, capture structure, scenario structure, and
fitted preprocessing can create dependencies between training and test data
~\cite{tavallaee2009kdd,sharafaldin2018cicids,garcia2014ctu13}.

% CR41: Condense the introductory protocol overview
\rev{We address these issues through a controlled intra-SNN evaluation. We screen
neuron--encoding combinations under a common architecture and training procedure,
then reevaluate them using a \emph{leakage-resistant confirmation protocol}.
This protocol fits transforms and categorical vocabularies on training data
only, fixes partitions independently of model seeds, preserves capture or
scenario structure where available, and audits duplication and class support.
These controls address identified train--test dependencies; other sources of
information sharing may remain.}

\begin{figure*}[!ht]
    \centering
    \includegraphics[width=\textwidth]{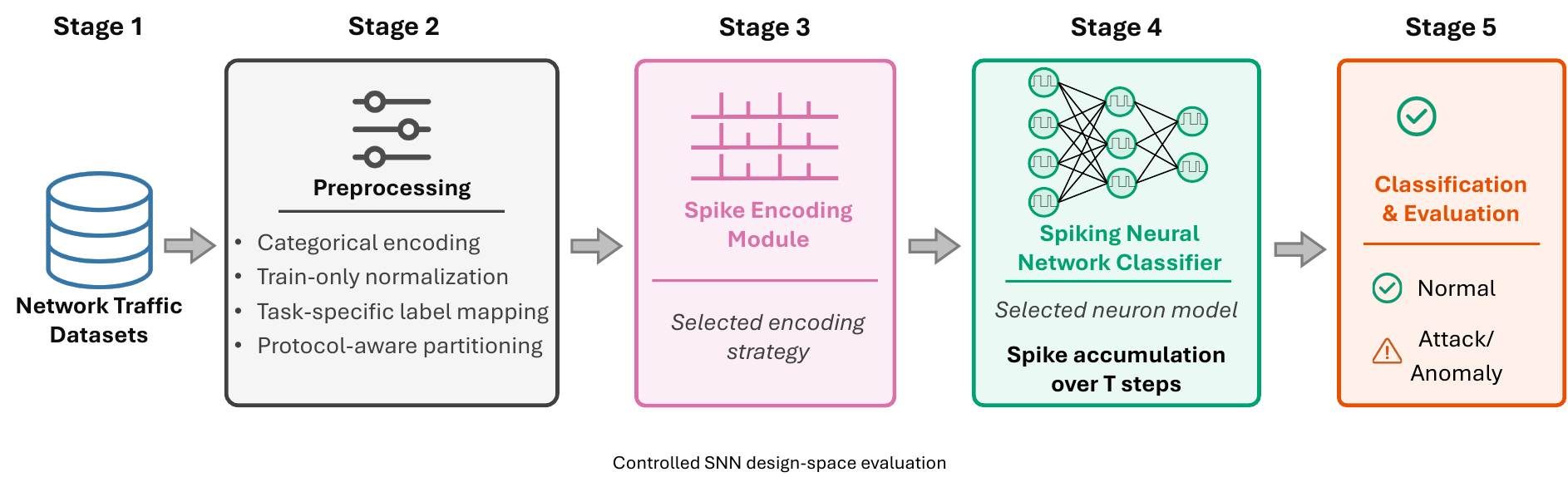}
    \caption{Experimental pipeline. Protocol-specific preprocessing and
    partitioning precede spike encoding, SNN inference over $T$ timesteps, and
    performance and efficiency evaluation. Only the encoding and neuron family
    vary in the $9\times3=\FNScrVariants$ screening sweep.}
    \label{fig:stages}
\end{figure*}

The design space contains nine neuron families implemented in
\texttt{snnTorch}~\cite{eshraghian2023training} and three spike encodings,
giving $9\times3=\FNScrVariants$ configurations. Figure~\ref{fig:stages}
summarizes the screening and confirmation pipeline used to evaluate these
choices.

Our scope is SNN design choice for static network-flow NIDS rather than general
SNN-versus-DNN superiority. We therefore examine neuron dynamics, spike
encoding, and evaluation protocol separately, particularly where temporal
structure is introduced by the model rather than observed in the data.

Based on this study design, we ask the following \textbf{Research Questions (RQs)}:

\begin{boxEnvRQ}
\begin{itemize}
\small

    \item \textbf{RQ1:} Which neuron and encoding choices remain promising when
    the SNN design space is reevaluated under a leakage-resistant confirmation
    protocol?

    \item \textbf{RQ2:} How do NIDS performance estimates change when
    train-only preprocessing and structure-aware partitioning are applied?

    \item \textbf{RQ3:} What does spike timing contribute to the neural
    representation, and how does input-code sparsity propagate into internal
    activity and SOPs?

\end{itemize}
\end{boxEnvRQ}

To address these questions, we make the following \textbf{contributions}:

\begin{boxEnvContrib}
\begin{itemize}
\small

    \item \textbf{C1:} We develop a controlled SNN design-space benchmark that
    evaluates nine neuron families and three spike encodings under a common
    architecture, training procedure, metric definition, and seed set.

    \item \textbf{C2:} We apply a leakage-resistant confirmation protocol using
    train-only transforms, seed-independent partitions, group-aware evaluation,
    causal CTU-13 host aggregation, duplicate accounting, and class-support
    auditing.

    \item \textbf{C3:} We construct matched-input controls that separate
    amplitude-to-time mapping from temporal spreading while preserving the
    active input feature set.

    \item \textbf{C4:} We decompose sparsity across input spikes, internal spike
    activity, and fanout-weighted SOPs to distinguish representation sparsity
    from computational activity.

    \item \textbf{C5:} We evaluate sensitivity to duplicate removal, scenario
    shift, unseen attack types, operating-point transfer, and
    screening-to-confirmation rank stability.

\end{itemize}
\end{boxEnvContrib}

%\noindent\textbf{Artifact:} \PH{ANONYMIZED ARTIFACT URL}

% =====================================================================
\section{Background and Related Work}\label{sec:related}

\subsection{Spiking Neurons and Readout}
Because the evaluated neuron families differ in how they maintain and update
temporal state, we use the leaky integrate-and-fire (LIF) model to establish a
common dynamical reference. For neuron $i$, the membrane state integrates
weighted presynaptic spikes and decays between timesteps,
\begin{equation}
    U_i[t] = \beta U_i[t-1] + \sum_j W_{ij}S_j[t] - \vartheta S_i[t-1],
\end{equation}
where $S_j[t]\in\{0,1\}$ denotes an incoming spike, $\beta$ is the
membrane-decay factor, and $\vartheta$ is the firing threshold. The neuron
emits
\begin{equation}
    S_i[t] = \Theta\!\left(U_i[t]-\vartheta\right),
\end{equation}
where $\Theta$ is the Heaviside step. Conductance-based families add a
synaptic state with decay $\alpha$, recurrent families feed previous spikes
back through learned connections, and gated families introduce additional
state and gating operations.

% CR48: Match the executable snnTorch 0.9.4 default ATan gradient.
% AUTHOR CHECK: This is surrogate.atan(alpha=2), including LeakyParallel's
% internal default. If the run code used custom scaling, report that formula.
Because the hard threshold is non-differentiable, we use the \rev{default}
arctangent surrogate gradient during backpropagation through time,
\begin{equation}
    \frac{\partial \widetilde{S}_i}{\partial U_i}
    = \rev{\frac{1}{1+\left(\pi(U_i-\vartheta)\right)^2}},
\end{equation}
following standard surrogate-gradient training
practice~\cite{eshraghian2023training}. We classify each sample from the
accumulated output-layer spike counts,
\begin{equation}
    \hat{y}=\arg\max_c \sum_{t=1}^{T} S_c[t],
\end{equation}
and optimize cross-entropy on those counts. Because membrane state decays
over time, two encodings containing the same number of spikes can produce
different hidden-state trajectories when the spikes occur at different
timesteps. This dependence on temporal placement motivates the matched-input
analysis in Section~\ref{sec:res-timing}.

\subsection{SNNs for Network Intrusion Detection} \label{sec:related-nids}
Prior work has established several ways of applying SNNs to network intrusion
and anomaly detection. Zhou and Li directly trained single-spike
temporal-coded SNNs on NSL-KDD and AWID and examined how leaky and non-leaky
neuron dynamics affect classification~\cite{zhou2021spiking}. Wang et al.
developed a convolutional SNN for intrusion classification using
CSE-CIC-IDS2018 and CIC-DDoS2019 traffic~\cite{wang2024efficient}. More
recently, Knapp et al. evaluated SNNs as the primary classifier, without
non-essential pre- or post-processing, across NSL-KDD, CIC-IDS2017,
CIC-IoT-2023, and AWID3~\cite{knapp2025efficacy}.

We also consider SNN-based anomaly-detection studies in related networked
settings, including automotive and industrial control environments. Jaoudi
et al. converted an unsupervised autoencoder to an SNN for detecting anomalous
CAN-bus traffic~\cite{jaoudi2020carhacking}, while Demertzis et al. used an
evolving one-class SNN for anomaly detection in industrial control and SCADA
data~\cite{demertzis2020gryphon}. These studies
demonstrate that SNNs can be applied across supervised intrusion
classification, one-class anomaly detection, and network-traffic monitoring.

The experimental choices across these studies differ in neuron model,
encoding, preprocessing, architecture, datasets, and task definition. As a
result, they do not provide a common basis for determining how neuron and
encoding choices affect NIDS performance independently. Our study addresses
this distinction by varying both design axes within one training and
evaluation pipeline, repeating the comparison under a leakage-resistant
confirmation protocol, and using matched-input controls to examine the
temporal representation without changing the active input feature set.

\subsection{Spike Encoding and Computational Activity}
When an SNN receives non-spiking observations, an encoding rule must convert
the input into spike events. Rate and temporal coding represent two common
families of such rules~\cite{auge2021encoding,yamazaki2022spiking}. Rate
coding represents a value through spike frequency or count over a time
window, whereas temporal coding uses the occurrence time of individual spikes
as part of the representation. Time-to-first-spike (TTFS) coding is a
temporal scheme in which an input value is mapped to the latency of its first
spike, with the particular value-to-time mapping determined by the encoding
rule~\cite{bonilla2022ttfs}.

This distinction is important for NIDS because the flow records considered in
this study are static feature vectors rather than native event streams. The
temporal axis is therefore introduced by the encoder. Changing the encoding
can alter both the information carried by spike timing and the number and
placement of events processed by the network. We consequently treat spike
encoding as a representation choice rather than assuming that temporal
structure in the resulting spike train was present in the original traffic
record.

Computational activity must also be distinguished from input-code sparsity.
A sparse encoder can generate few input spikes while recurrent membrane
dynamics produce additional internal activity. NeuroBench separates
hardware-independent algorithmic measurements from system-level measurements,
including quantities based on activation sparsity and synaptic
operations~\cite{yik2025neurobench}. Following that distinction, we report
input and internal spike counts together with fanout-weighted SOPs. SOPs characterize the amount of event-driven computation
executed by the model but are not interpreted as measurements of energy,
latency, or hardware efficiency.

\subsection{Evaluation Validity in Intrusion Detection}
Dataset construction and partitioning can materially affect
intrusion-detection results. Tavallaee et al. documented extensive redundancy
in KDDCup99 and introduced NSL-KDD to reduce several evaluation
artifacts~\cite{tavallaee2009kdd}. CIC-IDS2017 was constructed from multiple
capture periods and attack scenarios~\cite{sharafaldin2018cicids}, while
CTU-13 is organized around distinct botnet scenarios~\cite{garcia2014ctu13}.
These properties create dependencies that can be obscured when records are
randomly partitioned or when preprocessing is fitted before the train--test
boundary is established. They motivate the group-aware, order-aware, and
duplicate-sensitive confirmation procedures used in this study.

% =====================================================================
\section{Methodology}\label{sec:methodology}

% =====================================================================
\subsection{Study Design and Analysis Plan}\label{sec:scope}

RQ1 asks whether neuron and encoding choices remain supported when the
evaluation protocol changes. To avoid evaluating a preselected pairing, we
begin with the full $9\times3=\FNScrVariants$ factorial design defined by the
nine neuron families and three spike encodings in
Tables~\ref{tab:neurons} and~\ref{tab:encoders}. All configurations use the
same architecture and training procedure. Within each
$(\text{dataset},\text{seed})$ block, we rank configurations by macro-F1, and
the lowest mean rank identifies the screening target.

\begin{table}[!t]
\centering
% CR49: Distinguish dynamical order from the number of stored state tensors.
\caption{The nine \texttt{snnTorch} neuron families evaluated. ``Order''
\rev{denotes dynamical order or gated state structure}; ``Rec.'' marks explicit
recurrent feedback of output spikes.}
\label{tab:neurons}
\tabtiny
\begin{tabular}{@{}lccp{0.52\columnwidth}@{}}
\toprule
\textbf{Neuron} & \textbf{Order} & \textbf{Rec.}
& \textbf{Distinguishing mechanism} \\
\midrule
Alpha
& 2nd & --
& Alpha-function membrane response. \\

Lapicque
& 1st & --
& RC-circuit LIF formulation. \\

Leaky
& 1st & --
& LIF with membrane decay $\beta$. \\

LeakyParallel
& 1st & --
& Parallelized leaky implementation with an additional
$H$-dimensional bias vector. \\

RLeaky
& 1st & \checkmark
& LIF with learned recurrent spike feedback. \\

RSynaptic
& 2nd & \checkmark
& Synaptic LIF with recurrent spike feedback. \\

SConv2dLSTM
& gated & \checkmark
& Convolutional LSTM with spiking output. \\

SLSTM
& gated & \checkmark
& LSTM with threshold--spike--reset output. \\

Synaptic
& 2nd & --
& Two-state LIF with synaptic ($\alpha$) and membrane ($\beta$) decay. \\
\bottomrule
\end{tabular}
\tabnote{The two gated cells use firing threshold $0.1$ because the library
default produces no spikes on these inputs; all other families use the library
default. We fix $\beta=0.85$ and, where applicable, $\alpha=0.90$.}
\end{table}

\begin{table}[!t]
\centering
\caption{The three spike encoders evaluated. Each maps
$x\in[0,1]^d$ to spikes over $T$ timesteps. ``Spikes/feature'' is the expected
input-spike count for an active feature; ``Gate'' is the amplitude threshold
below which no spike is emitted.}
\label{tab:encoders}
\tabtiny
\begin{tabular}{@{}lccp{0.47\columnwidth}@{}}
\toprule
\textbf{Encoder} & \textbf{Spikes/feat.} & \textbf{Gate}
& \textbf{Mechanism} \\
\midrule
delta
& $1$
& $\theta_\delta=\FNDeltaThresh$
& Fires on positive transitions of a synthesized step signal holding
$x$ for $\lfloor T/2\rfloor$ steps; for static vectors this reduces to
one-step amplitude thresholding. \\

latency
& $1$
& $\theta_\mathrm{lat}=\FNLatThresh$
& Emits one spike per active feature at
% CR50: Use the same one-based timestep convention as the readout equations.
\rev{$t_i=1+\lfloor(1-x_i)(T-1)\rfloor$}; larger values fire earlier. \\

rate
& $T\,x_i$
& none
& Draws $S_{t,i}\sim\mathrm{Bernoulli}(x_i)$ at each timestep; magnitude is
represented by spike count. \\
\bottomrule
\end{tabular}
\tabnote{Latency and delta are sparse single-spike codes but differ in whether
feature magnitude is represented by spike time. Equalizing their gates makes
their active input spike sets identical for the matched-input controls.}
\end{table}

% CR42: Condense the two-axis study-design explanation
\rev{For RQ1, the leakage-resistant confirmation varies one factor at a time: the
encoding axis fixes the screening-selected neuron and compares all three
encodings, while the neuron axis fixes the selected encoding and compares all
nine neuron families. Changes in performance and ordering relative to screening address
RQ2.}

% CR43: Shorten the rationale for the complete sweep
\rev{To assess the joint ordering of all 27 configurations, we repeat}
% CR02: Clarify analysis-plan provenance
the complete design-space evaluation under confirmation. \rev{This sweep was added
after the primary two-axis results were known and documented in a separate
version-controlled addendum. We retain the two-axis analysis as primary and use
the complete sweep to assess screening-to-confirmation rank stability. The
plans were maintained in local version control, but the available records do
not independently establish pre-run commitment.}

% CR44: Condense the timing-analysis scope statement
\rev{For RQ3, matched-input timing controls examine temporal representation, while
spike and SOP measurements relate input sparsity to internal activity. Timing
decomposition and distribution-shift diagnostics are post hoc analyses used for
interpretation rather than configuration selection or confirmatory testing.}

% =====================================================================
\subsection{Datasets, Partitions, and Leakage-Resistant Confirmation}\label{sec:protocols}
To test whether the design choices generalize across the NIDS settings
discussed in Section~\ref{sec:related-nids}, we use NSL-KDD, KDDCup99,
CIC-IDS2017, and CTU-13. Their different traffic distributions and evaluation
structures provide the cross-dataset variation needed for RQ1 and RQ2.
Screening uses one partition per dataset, giving four blocks per seed.
Confirmation retains one
protocol each for NSL-KDD, KDDCup99, and CIC-IDS2017 and uses two
% CR20: Use primary-analysis terminology
preselected scenario-disjoint CTU-13 folds for the \rev{primary} axis
comparisons. The encoding axis therefore contains five confirmation protocols
per seed. The neuron axis uses four of these protocols because CIC-IDS2017
does not contain every class in its held-out five-class test interval.

NSL-KDD and KDDCup99 use a fixed five-class label set comprising Normal,
denial-of-service (DoS), Probe, remote-to-local (R2L), and user-to-root (U2R),
and their official train--test splits are retained~\cite{tavallaee2009kdd}.
Validation data for these two protocols are drawn as a stratified 10\% carve
from the official training pool using a fixed partition seed distinct from the
model seed. The model seed therefore changes weight initialization and batch
order but not the data partition. We verified this separation empirically:
two model seeds produce byte-identical validation- and test-set hashes on both
% CR03: Identify the implemented CIC-IDS2017 grouping
protocols, and the partition seed is recorded in the split manifest. CIC-IDS2017~\cite{sharafaldin2018cicids} \rev{uses a study-specific
five-class grouping documented in Table~\ref{tab:cic-labelmap}. The same
% CR71: Restrict the description to the evaluations reported in this paper.
label-conversion function is used for screening and confirmation.
Its KDD-style group names do not establish semantic
equivalence between the datasets' attack categories. For confirmation, we
construct a within-capture order-disjoint partition:} the first 70\% of each of eight capture files is
used for training, the next 10\% for validation, and the final 20\% for test
without shuffling. Because the redistributed data do not retain the original
\texttt{Timestamp}, row order is used only as a proxy for flow-writer ordering
and is not interpreted as chronological time.

CTU-13~\cite{garcia2014ctu13} is evaluated as a binary Normal-versus-Botnet
task. We construct four frozen scenario-disjoint folds and compute
per-source-host aggregate features causally from preceding flows of the same
source. Training, validation, and test sets contain complete, mutually
disjoint scenarios. Folds~0 and~1 are the two CTU-13 folds used in the
% CR21: Use primary-analysis terminology
\rev{primary} axis confirmation; all four folds are used separately to estimate
scenario-transfer performance. Scenario-disjoint evaluation does not imply
host-disjoint evaluation, and a source host can occur in different scenarios.

The leakage-resistant confirmation protocol is designed to reduce identifiable
dependencies between model development and held-out evaluation. We fit learned
transforms and categorical vocabularies using training rows only, fix
partitions independently of the optimization seed, preserve capture or
scenario structure when supported by the dataset, and audit each claimed
group separation. Unseen categorical values are assigned to a
training-defined \texttt{<unk>} level. For NSL-KDD, one such column is added
for each of its three categorical fields, changing the input dimension from
\FNDimNslScreen{} under screening to \FNDimNsl{} under confirmation.

We also measure exact train--test duplication instead of assuming that an
official split contains unique records. Duplicate-free sensitivity analyses
are performed where overlap is present. Table~\ref{tab:audit} summarizes the
partition and duplication audit.

\begin{table}[!t]
\centering
\caption{Split audit for the confirmation protocols. ``Shared'' counts
overlapping group identifiers and is a violation only when that group is
% CR22: Use primary-analysis terminology
claimed disjoint. Bullets mark the \rev{primary} CTU-13 confirmation folds.}
\label{tab:audit}
\tabtiny
\begin{tabular}{@{}llp{0.40\columnwidth}rr@{}}
\toprule
\textbf{Dataset} & \textbf{Group} & \textbf{Claimed separation}
& \textbf{Shared} & \textbf{Dup.\ (\%)} \\
\midrule
NSL-KDD     & --       & Official train/test           & 0 & \FNDupNsl \\
KDDCup99    & --       & Official train/test           & 0 & \FNDupKdd \\
CIC-IDS2017 & Capture  & Within-capture order-disjoint & 8 & 10.86 \\
CTU-13      & Scenario & Scenario-disjoint, fold 0 $\bullet$ & 0 & 0.00 \\
CTU-13      & Scenario & Scenario-disjoint, fold 1 $\bullet$ & 0 & 0.00 \\
CTU-13      & Scenario & Scenario-disjoint, fold 2 & 0 & 0.00 \\
CTU-13      & Scenario & Scenario-disjoint, fold 3 & 0 & 0.00 \\
\bottomrule
\end{tabular}
\end{table}

Table~\ref{tab:ctufolds} records the CTU-13 scenario assignments and sample
counts. Performance associated with these partitions is reported with the
experimental results rather than used to define the folds.

\begin{table}[!t]
\centering
\caption{CTU-13 scenario-disjoint partitions. Validation and test sets contain
complete scenarios disjoint from training; bullets mark the two folds used in
% CR23: Use primary-analysis terminology
the \rev{primary} axis confirmation.}
\label{tab:ctufolds}
\tabtiny
\begin{tabular}{@{}lllrrr@{}}
\toprule
\textbf{Fold} & \textbf{Test sc.} & \textbf{Val. sc.}
& \textbf{Train} & \textbf{Val.} & \textbf{Test} \\
\midrule
0 $\bullet$ & 2,4,8    & 1,3 & 378{,}980 & 57{,}909 & 364{,}243 \\
1 $\bullet$ & 7,9,11   & 1,3 & 532{,}279 & 57{,}909 & 210{,}944 \\
2           & 1,3,5,10 & 0   & 648{,}869 & 71{,}348 & 80{,}915  \\
3           & 0,6,12   & 1,3 & 598{,}193 & 57{,}909 & 145{,}030 \\
\bottomrule
\end{tabular}
\end{table}

% =====================================================================
\subsection{SNN Design Space and Training} \label{sec:enc}
Each encoder maps the normalized feature vector $x\in[0,1]^d$ to a spike
representation over $T$ timesteps. Because the latency and delta encoders use
different amplitude gates, their direct comparison can mix temporal
representation with differences in which input features are allowed to spike.
We therefore measure the feature mass removed by each gate and evaluate two
gate-matched controls: one removes the latency gate, and the other sets it equal
to the delta gate. These controls allow us to determine whether an observed
encoding difference is associated with temporal representation or with the
% CR04: Point to retained-feature counts
active input feature set.
\rev{Tables~\ref{tab:sparsity}--\ref{tab:gatecontrol} report the mean active
features retained per sample. For the single-spike encoders, these equal the
input-spike counts; gating does not reduce input dimensionality.}

All models use a two-layer spiking multilayer perceptron with hidden width
$H=128$ and simulation length $T=25$. We use membrane decay $\beta=0.85$,
synaptic decay $\alpha=0.90$ where applicable, and the arctangent surrogate
gradient. Training uses Adam with learning rate $10^{-3}$, batch size $128$,
% CR05: Scope the five-seed statement
and 10 epochs. \rev{Each configuration in the main screening and
confirmation benchmarks is evaluated over seeds $\{42,43,44,45,46\}$} on a single NVIDIA H200 GPU using PyTorch~2.12 and
\texttt{snnTorch}~0.9.4.

We do not tune hyperparameters separately for individual neuron families. The
neuron comparison therefore measures performance under a shared operating
point rather than each family's optimized potential. The gated
\texttt{SLSTM} and \texttt{SConv2dLSTM} cells use threshold $0.1$ because
the library default of $1.0$ produces no spikes on these sparse tabular
inputs. This operating-point constraint is considered when interpreting the
neuron comparison.

% =====================================================================
\subsection{Metrics and Statistical Analysis} \label{sec:metrics}
Because the datasets are class-imbalanced and contain rare attacks, macro-F1
is the primary metric so that each class contributes equally to configuration
selection. Accuracy, MCC, DR, FAR, and per-class metrics provide complementary
views of overall correctness, attack detection, false alarms, and
class-specific failure modes. Macro-F1 is computed over the fixed task label
set, so a class with no test support contributes zero. Differences between
percentage-valued metrics are reported in percentage points (pp).

For screening, each $(\text{dataset},\text{seed})$ pair forms one paired
block, giving $4\times5=20$ blocks. Because macro-F1 occurs at different
levels across datasets, we rank the 27 configurations within each block and
use mean rank across blocks as the cross-dataset selection statistic. The
configuration with the lowest mean rank becomes the screening target. We
report mean $\pm$ standard deviation across seeds within each dataset rather
than pooling cross-dataset standard deviations.

We use the Friedman test to assess whether configuration ranks differ across
the full design space. When it rejects, we apply the Nemenyi post hoc
procedure and report its critical difference (CD), following standard
multiple-classifier comparison practice~\cite{demsar2006statistical}. The CD
describes the resolution of the all-pairs comparison; failure to separate two
configurations is not interpreted as evidence of equivalence.

We also specify focused paired comparisons between the screening target and
its same-neuron encoding competitor and between the target and the leading
configuration using another neuron family. We evaluate these comparisons with
paired Wilcoxon signed-rank tests~\cite{wilcoxon1945individual} and control
multiplicity using the Holm procedure~\cite{holm1979simple}. The accompanying
hierarchical bootstrap resamples datasets first and seeds within datasets
second, so repeated seeds are not treated as independent datasets.
% AUTHOR CHECK: Add the bootstrap confidence level, resample count, and RNG
% seed from the saved analysis record. These settings were not supplied.

The encoding-axis confirmation evaluates all three encodings while holding the
screening-selected neuron fixed across five confirmation protocols
(\FNEncBlocks{} protocol--seed blocks). The neuron-axis confirmation evaluates
all nine neuron families while holding the screening-selected encoding fixed
across four protocols (\FNNeuBlocks{} blocks). CIC-IDS2017 is omitted from
the neuron axis because one class is absent from its held-out five-class test
interval.

% CR51: State the separately frozen permutation counts for both tests.
\rev{For the complete confirmation sweep, we rank all 27 configurations over
the five confirmation protocols and compare this ordering with screening.
We assess agreement using Kendall's rank correlation coefficient
~\cite{kendall1938rank}, with \FNTauPerms{} permutations for the full ranking
and \FNTauTopPerms{} for the configurations within one screening CD of the
leader, together with top-weighted rank measures. These analyses assess
whether the design-selection ordering persists under confirmation.}
% AUTHOR CHECK: Verify the permutation-tail and Monte Carlo correction
% conventions against the saved test records; counts above follow frozen macros.

A post-run metric audit identified an implementation in which the macro-F1
denominator could depend on the set of predicted classes. We recomputed the
affected values from persisted predictions using the fixed task label set,
without retraining. The correction changed one reported value but did not
change configuration rankings or hypothesis-test conclusions; the full audit
is provided in the appendix.

% =====================================================================
\subsection{Timing and Computational Activity Analyses}
\label{sec:measurements}
\label{sec:eff}

For each saved model, we evaluate the classifier after every timestep,
\begin{equation}
    \hat{y}(t)=\arg\max_c\sum_{\tau=1}^{t}S_c(\tau),
\end{equation}
and record macro-F1, MCC, DR, and cumulative spike activity. We define
$T_{95}$ and $T_{99}$ as the first timesteps at which macro-F1 reaches
95\% and 99\% of its terminal value.

To separate two roles of spike timing, we compare three matched-input arms.
The \emph{mapped} arm uses latency coding. The \emph{spread-only} arm
preserves the same active features and per-timestep spike counts while
permuting feature-to-time assignments. The \emph{no-spread} arm places the
same active-feature spike set at one timestep. We define
\begin{equation}
    \dtime=\dmap+\dspread,
\end{equation}
where $\dmap$ is the mapped-minus-spread-only difference and $\dspread$ is
the spread-only-minus-no-spread difference. We repeat the decomposition for
$T\in\{5,10,25,50\}$ and verify the identity at every shared
% CR06: Document the three-seed analysis provenance
protocol--$T$ cell. This decomposition is used only for post hoc mechanism
analysis.

\rev{The gate controls, matched-input timing decomposition, and auxiliary
timestep sweeps use seeds 42--44. At $T=25$, the
mapped-latency arm reuses the corresponding benchmark test predictions without
retraining; its three-seed mean therefore differs from the five-seed mean in
Table~\ref{tab:strict-results}. At $T\in\{5,10,50\}$, the mapped models are
trained separately at each simulation length. Across these analyses, the
confirmation partitions have identical hashes, and training-only preprocessing
and shared training settings are unchanged; only the designated encoding,
amplitude gate, or simulation length varies.}

We separately measure input, hidden, and output spike activity. To account for
the computation driven by each event, we report fanout-weighted synaptic
operations (SOPs),
\begin{equation}
    \mathrm{SOP}
    = \sum_{\ell,t}s_{\ell}(t)\,\mathrm{fanout}_{\ell}.
\end{equation}
Gated cells are charged four input projections to represent their gate
operations. Following hardware-independent neuromorphic benchmarking
practice~\cite{yik2025neurobench}, SOPs are treated as operation counts rather
than measurements of energy or deployment latency. Output-layer spikes have
zero fanout because they drive no downstream projection.

\section{Results and Discussion}\label{sec:results}

\subsection{Design-space screening}\label{sec:res-rank}

\emph{(RQ1, screening)} Across the \FNScrVariants{} configurations and
\FNScrBlocks{} paired dataset--seed blocks, \texttt{LeakyParallel/latency}
attains the lowest mean macro-F1 rank (\FNScrRank). The Friedman test rejects
equal performance ($\chi^2=\FNScrChi$, $p=\pv{\FNScrP}$), with a Nemenyi
critical difference of \FNScrCD. The rank-1 configuration is therefore not
statistically isolated: \FNScrWithinCD{} of the \FNScrVariants{} configurations
lie within one critical difference, and \texttt{Leaky/latency} is not separated
from \texttt{LeakyParallel/latency} ($p=\FNScrLeakyP$).

The focused comparison with \texttt{LeakyParallel/rate} favors latency in
\FNScrWinsA{} of \FNScrBlocks{} blocks, with a median paired difference of
\FNScrMedDelta{}~pp macro-F1 (Holm-adjusted $p=\FNScrHolmP$). The hierarchical
bootstrap gives $[\FNScrBootLo,\FNScrBootHi]$~pp. At the encoding level, the
block-level comparison also favors latency ($p=\FNScrEncP$), whereas the
dataset-level bootstrap interval $[\FNScrEncCiLo,\FNScrEncCiHi]$ includes zero.
The screening evidence therefore supports an aggregate latency preference, but
not a uniform effect across datasets.

% CR24: Use primary-analysis terminology
KDDCup99 illustrates why the \rev{primary metric} matters. Rate has
slightly higher accuracy (\FNKddAccRat{} versus \FNKddAccLat{}), whereas latency
has the higher macro-F1 by \FNKddFOneGain{}~pp. Removing CTU-13 from the
screening aggregation also exchanges the first two configurations:
\texttt{Leaky/latency} has mean rank \FNNoCtuLeadRank{} and
\texttt{LeakyParallel/latency} \FNNoCtuChampRank{}, a difference of only
\FNNoCtuMargin{} rank units. Latency-coded configurations still occupy all
\FNNoCtuLatTop{} leading positions. We therefore treat
\texttt{LeakyParallel/latency} and \texttt{Leaky/latency} as a co-leading pair
rather than inferring a resolved preference between them.

% CR45: Tighten the Leaky implementation comparison
% CR52: Count only trainable parameters, excluding frozen state matrices.
\rev{\texttt{LeakyParallel} parallelizes the \texttt{Leaky} formulation and adds an
$H$-dimensional bias vector. For NSL-KDD at confirmation dimension $d=125$,
their \rev{trainable parameter counts} are 16{,}901 and 16{,}773, respectively; at screening
dimension $d=122$, they are 16{,}517 and 16{,}389. The 128-parameter
difference is unchanged. Their statistical non-separation is consistent with
closely related implementations; the small screening rank gap does not
establish a difference in underlying neuron dynamics.}

\subsection{Leakage-resistant confirmation}\label{sec:res-strict}
\emph{(RQ1, confirmation)} The confirmation results preserve the aggregate
encoding preference while revealing a protocol-specific inversion and a less
concentrated neuron-family ordering.

\paragraph{Encoding axis}
Across \FNEncBlocks{} paired protocol--seed blocks, latency has the lowest mean
rank (\FNEncRank{} of \FNEncLevels{} encodings, Friedman
$p=\pv{\FNEncP}$) and ranks first in \FNEncWins{} blocks. It has the highest
terminal macro-F1 on four of the five protocols in
Table~\ref{tab:strict-results}. On those four protocols, latency also has the
highest MCC. Its DR is highest on four of the five protocols; on KDDCup99, rate has DR
$0.9129$ versus $0.9122$ for latency while latency retains the higher macro-F1
and MCC. FAR does not favor one encoding uniformly: delta has the lowest FAR on
NSL-KDD, KDDCup99, and CTU-13 fold~1, whereas latency has the lowest FAR on
CIC-IDS2017 and CTU-13 fold~0.

CTU-13 fold~1 reverses the macro-F1 ordering. Delta reaches $0.8825$ macro-F1
and $0.7668$ MCC, compared with $0.7750/0.6251$ for rate and
$0.7702/0.6220$ for latency. Its FAR is also lower ($0.1601$ versus $0.4440$
and $0.4562$), although its DR decreases to $0.9213$ from $0.9928$ and
$0.9968$. The inversion is therefore an operating trade-off across macro-F1,
MCC, DR, and FAR rather than a change in macro-F1 alone. SOP counts provide a
second distinction: delta has the lowest SOP count on all five protocols, rate
the highest, and latency lies between them. Only on CTU-13 fold~1 does delta's
lowest SOP count coincide with the highest macro-F1.

\begin{table*}[!ht]
\centering
% CR32/CR70: Predictive metrics include SDs; SOPs are five-seed means.
\caption{Encoding-axis confirmation with \texttt{LeakyParallel} fixed.
\rev{Predictive metrics are mean $\pm$ across-seed SD over five seeds.}
Bold marks the highest macro-F1 within each protocol; SOPs are operation counts.}
\label{tab:strict-results}
\tabsmall
\scriptsize
\begin{tabular}{@{}llccccr@{}}
\toprule
\textbf{Protocol} & \textbf{Configuration} & \textbf{Macro-F1} & \textbf{MCC} &
\textbf{DR} & \textbf{FAR} & \textbf{SOPs/sample} \\
\midrule
\multirow{3}{*}{NSL-KDD}
 & \textbf{latency} & $\mathbf{0.5846 \pm 0.0279}$ & $0.6585 \pm 0.0272$ & $0.6696 \pm 0.0299$ & $0.0648 \pm 0.0145$ & 3{,}363 \\
 & rate             & $0.5471 \pm 0.0293$ & $0.6494 \pm 0.0154$ & $0.6536 \pm 0.0197$ & $0.0527 \pm 0.0217$ & 32{,}502 \\
 & delta            & $0.5197 \pm 0.0160$ & $0.6417 \pm 0.0148$ & $0.6488 \pm 0.0207$ & $0.0287 \pm 0.0004$ & 2{,}421 \\
\midrule
\multirow{3}{*}{KDDCup99}
 & \textbf{latency} & $\mathbf{0.6309 \pm 0.0119}$ & $0.8269 \pm 0.0032$ & $0.9122 \pm 0.0010$ & $0.0130 \pm 0.0049$ & 2{,}843 \\
 & rate             & $0.6089 \pm 0.0400$ & $0.8266 \pm 0.0033$ & $0.9129 \pm 0.0010$ & $0.0125 \pm 0.0045$ & 33{,}225 \\
 & delta            & $0.6083 \pm 0.0365$ & $0.8207 \pm 0.0023$ & $0.9119 \pm 0.0008$ & $0.0076 \pm 0.0003$ & 1{,}920 \\
\midrule
\multirow{3}{*}{CIC-IDS2017}
 & \textbf{latency} & $\mathbf{0.7842 \pm 0.0030}$ & $0.9812 \pm 0.0033$ & $0.9745 \pm 0.0038$ & $0.0006 \pm 0.0003$ & 6{,}817 \\
 & rate             & $0.7277 \pm 0.0153$ & $0.9431 \pm 0.0050$ & $0.9264 \pm 0.0076$ & $0.0024 \pm 0.0004$ & 48{,}834 \\
 & delta            & $0.4981 \pm 0.0087$ & $0.7680 \pm 0.0094$ & $0.7102 \pm 0.0487$ & $0.0095 \pm 0.0044$ & 5{,}263 \\
\midrule
\multirow{3}{*}{CTU-13 f0}
 & \textbf{latency} & $\mathbf{0.9985 \pm 0.0013}$ & $0.9971 \pm 0.0026$ & $0.9981 \pm 0.0024$ & $0.0008 \pm 0.0009$ & 3{,}399 \\
 & rate             & $0.9876 \pm 0.0018$ & $0.9754 \pm 0.0035$ & $0.9804 \pm 0.0033$ & $0.0016 \pm 0.0005$ & 40{,}195 \\
 & delta            & $0.5938 \pm 0.0048$ & $0.3461 \pm 0.0057$ & $0.3748 \pm 0.0072$ & $0.0713 \pm 0.0005$ & 2{,}930 \\
\midrule
\multirow{3}{*}{CTU-13 f1}
 & latency          & $0.7702 \pm 0.0041$ & $0.6220 \pm 0.0046$ & $0.9968 \pm 0.0025$ & $0.4562 \pm 0.0088$ & 3{,}084 \\
% CR07: Standardize the CTU-13 rate cell precision
  & rate             & \rev{$\FNCtuBEncRat \pm 0.0050$} & $0.6251 \pm 0.0070$ & $0.9928 \pm 0.0005$ & $0.4440 \pm 0.0089$ & 40{,}085 \\
 & \textbf{delta}   & $\mathbf{0.8825 \pm 0.0067}$ & $\mathbf{0.7668 \pm 0.0142}$ & $0.9213 \pm 0.0130$ & $\mathbf{0.1601 \pm 0.0007}$ & \textbf{2{,}806} \\
\bottomrule
\end{tabular}
% CR08: Distinguish input determinism from model activity
\tabnote{The latency and delta encoders are deterministic given $x$, so their
\rev{input spike counts} do not vary across seeds; only the rate encoder redraws per seed.
% CR70: The repo audit confirms these are means over all five benchmark seeds.
\rev{SOPs are means over five seeds, as in Table~\ref{tab:sopdecomp};
internal activity can vary across trained seeds.} The
spread-only permutation used in the timing controls is keyed on a separate
control seed and varies per seed.}
\end{table*}

\paragraph{Neuron axis}
Across \FNNeuBlocks{} blocks, \texttt{LeakyParallel} has the lowest aggregate
mean rank (\FNNeuRank{} of \FNNeuLevels{} families, Friedman
$p=\pv{\FNNeuP}$), followed by \FNNeuRunnerUp{} at
\FNNeuRunnerUpRank. The ordering is not a per-block dominance result:
\FNNeuMostWins{} wins more individual blocks than \texttt{LeakyParallel}, and
on CTU-13 fold~1 \texttt{RSynaptic} and \texttt{RLeaky} reach macro-F1 values
of \FNCtuBNeuRSyn{} and \FNCtuBNeuRLky{}, respectively, compared with
\FNCtuBNeuLP{} for \texttt{LeakyParallel}. Table~\ref{tab:neuron-results}
reports the full per-protocol metrics. We interpret these ranks only at the
shared operating point. In particular, CTU-13 fold~1 contains clear optimization
failures: \texttt{SLSTM} has MCC $-0.3332$, while \texttt{SConv2dLSTM} has MCC
$0.1895\pm0.5524$. Those cells are not used to infer intrinsic neuron-family
quality.

\paragraph{Complete design-space confirmation}\label{sec:res-full}
The separate complete confirmation sweep gives the same rank-1 configuration:
\texttt{LeakyParallel/latency} has mean rank \FNConfRank{}, and the
\FNConfLatTop{} leading positions are latency-coded. Friedman again rejects
equal performance ($\chi^2=\FNConfChi$, $p=\pv{\FNConfP}$). The Nemenyi
critical difference is \FNConfCD{} for \FNConfBlocks{} blocks, compared with
\FNScrCD{} under screening. This resolution does not justify treating
non-separated configurations as equivalent.

When the complete ranking is grouped by encoding, latency ranks first in
\FNConfEncWinsLat{} of \FNConfBlocks{} blocks, compared with
% CR09: Disambiguate the rank scale
\FNConfEncWinsRat{} for rate and \FNConfEncWinsDel{} for delta; \rev{the
corresponding mean configuration ranks are \FNConfEncLat, \FNConfEncRat{}, and
\FNConfEncDel{}, respectively. These averages remain on the 27-configuration
scale, rather than the three-encoding scale.} The
three non-latency block wins comprise two NSL-KDD blocks and one CTU-13 fold~1
block.

For the two complete 27-configuration rankings, Kendall's rank correlation is
computed from concordant and discordant configuration pairs,
\begin{equation}
    \tau=\frac{C-D}{\binom{n}{2}},
\end{equation}
where $C$ is the number of configuration pairs ordered in the same direction
under screening and confirmation, $D$ is the number ordered in opposite
directions, and $n=27$. Among the $\binom{27}{2}=351$ configuration pairs,
317 are concordant and 34 are discordant, yielding Kendall
$\tau=\FNTau$ (permutation $p\,\FNTauP$). Thus, most pairwise ordering
relationships identified during screening are preserved under confirmation.
Within the \FNTauTopN{} configurations inside one screening CD,
$\tau=\FNTauTop$, compared with \FNTauBottom{} over the remaining
\FNTauBottomN{}, showing that the agreement is not confined to the
lower-ranked tail. Top-weighted measures provide the same interpretation:
% CR53: Name the finite-depth RBO variant that reproduces the frozen values.
RBO~\cite{webber2010rbo}, \rev{truncated at depth \FNConfVariants{} without tail
extrapolation}, is \FNRBOninety{} at persistence $p=0.9$ and
\FNRBOeighty{} at $p=0.8$; AP rank correlation~\cite{yilmaz2008ap} is
\FNTauAPscr{} with screening as the reference and \FNTauAPconf{} with
confirmation as the reference; and top-$k$ membership overlaps are
\FNTopKthree{}/3, \FNTopKfive{}/5, \FNTopKten{}/10, and
\FNTopKfourteen{}/14. The leading-subset permutation test gives
$p=\FNTauTopP$.

% CR46: Condense the rank-movement interpretation
\rev{\texttt{SConv2dLSTM/delta} gains seven positions,
\texttt{Lapicque/latency} loses seven, and \texttt{LeakyParallel/latency}
remains first. Table~\ref{tab:conf27} and Fig.~\ref{fig:rank-shift} detail these
movements. The agreement supports screening for design selection, with
confirmation providing the final performance estimates and comparisons.}

\subsection{Protocol sensitivity and boundary conditions}\label{sec:res-limits}
\paragraph{CTU-13 protocol effect}
\emph{(RQ2)} CTU-13 exhibits the largest change when the evaluation boundary is
made stricter. \texttt{LeakyParallel/latency} has \FNCtuLegacy\% macro-F1 with
whole-scenario aggregation and a random flow split, \FNCtuCausalRandom\% after
making the aggregation causal, and \FNCtuStrict\% after scenario-disjoint
partitioning. The causal-aggregation change accounts for \FNCtuCausalCost{}~pp,
whereas the scenario-disjoint step accounts for \FNCtuScenarioCost{}~pp, for a
total reduction of \FNCtuTotalCost{}~pp. Table~\ref{tab:ctu-fold-performance} 
reports the four fold means. Their
four-fold aggregate is \FNCtuStrict\%, with SD \FNCtuSdEval{}~pp over
\FNCtuNEval{} fold--seed evaluations. The
\FNCtuIdxFB\%--\FNCtuIdxFD\% range shows that scenario difficulty is hidden by
the screening partition and is not summarized adequately by the aggregate
alone.

\begin{table}[!t]
\centering
\caption{CTU-13 scenario-transfer macro-F1 by fold for
\texttt{LeakyParallel/latency}, five seeds. Bullets mark the two folds used in
% CR25: Use primary-analysis terminology
the \rev{primary} confirmation analysis.}
\label{tab:ctu-fold-performance}
\tabtiny
\begin{tabular}{@{}lc@{}}
\toprule
\textbf{Fold} & \textbf{Macro-F1 (\%)} \\
\midrule
0 $\bullet$ & \FNCtuIdxFA \\
1 $\bullet$ & \FNCtuIdxFB \\
2           & \FNCtuIdxFC \\
3           & \FNCtuIdxFD \\
\midrule
Four-fold mean & \FNCtuStrict \\
\bottomrule
\end{tabular}
\tabnote{The SD over the \FNCtuNEval{} fold--seed evaluations is
\FNCtuSdEval{}~pp. Per-fold standard deviations are not reported here.}
\end{table}

\paragraph{Duplicate and near-tie sensitivity}
Exact train--test overlap is large enough to warrant an explicit check:
\FNDupKdd\% of the official KDDCup99 test set and \FNDupNsl\% of NSL-KDD occur
verbatim in training. Removing duplicates preserves the encoding leader and the
complete encoding ordering on all \FNDedupProtocols{} audited protocols
(Table~\ref{tab:dedup}). On KDDCup99, the median paired latency--rate margin is
\FNDedupKddOffMedian{}~pp on the official test set and
\FNDedupKddDedMedian{}~pp after duplicate removal. Latency is ahead in only
three of the five seed blocks, so the positive median is not interpreted as a
uniform per-seed effect. The neuron-family leader is also unchanged, although
the complete neuron ordering changes on KDDCup99. We therefore do not attribute
the encoding ordering to memorized duplicate records.

One KDDCup99 comparison remains too close to resolve. Under the screening
protocol, the rate--delta ordering rests on a mean macro-F1 margin of
\FNKddRateDeltaMarginScr{}~pp, with per-seed differences spanning
\FNKddRateDeltaLoScr{} to \FNKddRateDeltaHiScr{}~pp; delta is ahead in
\FNKddDeltaWinsScr{} seed blocks and rate in \FNKddRateWinsScr{}. Under
confirmation, Table~\ref{tab:strict-results} reports mean macro-F1 of $0.6089$
for rate and $0.6083$ for delta, corresponding to a margin of
\FNKddRateDeltaMarginConf{}~pp. Both evaluations therefore place this pair near
the decision boundary, and the protocol-level ordering is treated
descriptively rather than as a reliable preference.

The KDD-family latency effect is also concentrated in the rare classes. Removing
R2L and U2R from the descriptive macro-F1 reverses the latency--rate ordering by
only $0.09$~pp on NSL-KDD and $0.29$~pp on KDDCup99.
% CR54/CR68: Use the author's repo-audited class-variance shares.
% The denominator is the sum of across-seed class-F1 variances, not the
% variance of macro-F1 (which also contains cross-class covariance terms).
% Values supplied from the committed-repo audit: U2R 74.7%, R2L 19.6%.
\rev{For \texttt{LeakyParallel/latency} on NSL-KDD, U2R and R2L account for
74.7\% and 19.6\%, respectively, of the summed across-seed class-F1
variances (Table~\ref{tab:perclass}).}
These checks localize much of the latency margin to the rare-class
% CR10: Qualify rare-class ranking precision
portion of the task rather than to a broad accuracy advantage.
\rev{Small differences among leading configurations should therefore be
interpreted cautiously; screening-to-confirmation rank agreement does not
establish robustness to alternative samples of rare-class examples.}

\paragraph{CIC-IDS2017 class support}
% CR11: Explain the CIC-IDS2017 support boundary
\rev{The mapped U2R group is present in CIC-IDS2017 training and validation
but absent from the order-disjoint confirmation test interval. Web-attack
records occur only in training, whereas all Heartbleed records occur in test;
Heartbleed is thus an unseen native attack type within the supported R2L group.}
We retain
the fixed five-class macro-F1 for ranking and report the test-present-class
value only descriptively. For \texttt{LeakyParallel/latency}, the fixed five-class macro-F1 is
\FNCicFixedLat{}, while the test-present-class value is
\FNCicPresentLat{}, computed from the four supported class F1 values in
Table~\ref{tab:perclass}. Because the unsupported class contributes the same
zero to every configuration, this rescaling does not change the within-protocol
ordering. The fixed five-class score is capped at \FNCicCeiling{} on this
protocol and should not be interpreted as evidence that CIC-IDS2017 is
intrinsically more difficult than the other datasets.

\paragraph{Operating-point transfer and unseen attacks}
Table~\ref{tab:far} in Appendix~\ref{app:boundary} shows that the
validation-calibrated FAR transfers differently across protocols. Validation-to-test FAR changes from
\FNFarValCic\% to \FNFarTestCic\% on CIC-IDS2017, \FNFarValKdd\% to
\FNFarTestKdd\% on KDDCup99, \FNFarValNsl\% to \FNFarTestNsl\% on NSL-KDD,
\FNFarValCtuA\% to \FNFarTestCtuA\% on CTU-13 fold~0, and \FNFarValCtuB\% to
\FNFarTestCtuB\% on fold~1. DR at the same frozen thresholds is
\FNFarDrCic\%, \FNFarDrKdd\%, \FNFarDrNsl\%, \FNFarDrCtuA\%, and
\FNFarDrCtuB\%, respectively. The CTU-13 fold~1 increase to
\FNFarTestCtuB\% while retaining \FNFarDrCtuB\% DR identifies a calibration
% CR12: State the operational implication of FAR
transfer problem, rather than a change in the argmax results of
Table~\ref{tab:strict-results}.
\rev{Both fold~1 operating points flag nearly half of benign traffic. Their
high detection rates therefore do not establish deployment readiness; macro-F1
and DR must be interpreted alongside FAR under the relevant decision rule.}

Separately, \FNUnseenFrac\% of KDDTest+ attack rows belong to attack types
absent from training. Across all \FNUnseenConfigs{} configurations, DR falls
from \FNKnownLo--\FNKnownHi{} on known types to \FNUnseenLo--\FNUnseenHi{} on
unseen types. Because this degradation occurs across the design space, we treat
it as a task-level generalization boundary rather than an encoding- or
neuron-specific result.

\subsection{Temporal convergence and computational activity}\label{sec:res-activity}

\begin{figure*}[!t]
  \centering
  \includegraphics[width=0.85\textwidth]{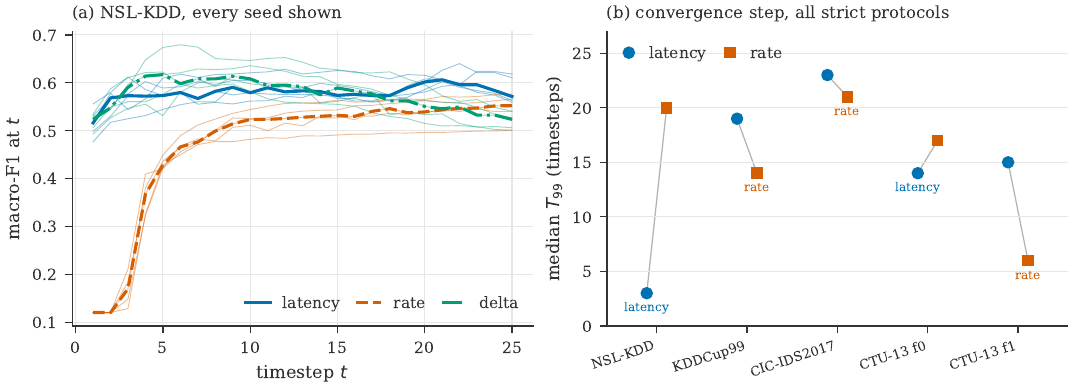}
  \caption{Temporal evidence accumulation for \texttt{LeakyParallel}. NSL-KDD
  shows all seeds; across protocols, latency reaches $T_{99}$ earlier on two and
  rate on three.}
  \label{fig:readout}
\end{figure*}

\paragraph{Temporal convergence}
On NSL-KDD, median $T_{99}$ is \FNTNNNslLat{} for latency and
\FNTNNNslRat{} for rate. Across the five confirmation protocols, however,
latency reaches $T_{99}$ earlier on only two and rate on three
(Table~\ref{tab:readout}, Fig.~\ref{fig:readout}).
% CR55: Separate observed readout convergence from guaranteed input delivery.
\rev{Delta has median $T_{99}=1$ on NSL-KDD. Its static-input encoder places
all input spikes at the first timestep, although subsequent neural dynamics
can still change predictions.} It nonetheless finishes \FNDeltaDeficitNsl{}~pp below latency at the terminal
timestep. $T_{99}$ therefore measures convergence to a configuration's own
terminal performance, not the quality of that terminal performance.

\paragraph{Spike activity and SOPs}
\emph{(RQ3, activity)} Rate coding produces \FNInpRatioMin$\times$--\FNInpRatioMax$\times$
more input spikes than latency across the confirmation protocols, whereas total
network spikes differ by only \FNTotRatioMin$\times$--\FNTotRatioMax$\times$.
The input-level difference is therefore not propagated one-for-one through the
network. It still matters computationally because it occurs before the widest
projection: rate requires \FNSopRatioMin$\times$--\FNSopRatioMax$\times$ more
SOPs per sample than latency. Tables~\ref{tab:input-spikes} and
\ref{tab:sopdecomp} report the corresponding activity and layer-level operation
counts.

\begin{figure}[!t]
\centering
% CR56: Bundled vector figure has the annotation corrected to an SOP comparison.
\includegraphics[width=0.85\columnwidth]{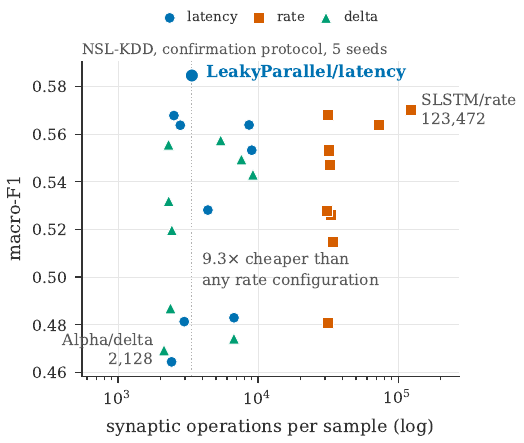}
\caption{Macro-F1 versus SOPs for all \FNScrVariants{} confirmation
configurations on NSL-KDD. \texttt{LeakyParallel/latency} is Pareto-optimal;
the lowest-SOP rate-coded configuration requires 9.3$\times$ as many SOPs.}
\label{fig:pareto27}
\end{figure}

The full confirmation design space shows the same separation between quality
and operation count. On NSL-KDD, SOP count spans
\FNParetoCostSpan$\times$ across all \FNScrVariants{} configurations, and the
lowest-SOP rate-coded configuration still requires 9.3$\times$ as many SOPs as
\texttt{LeakyParallel/latency}. The latter is also Pareto-optimal in the
standard multi-objective sense~\cite{miettinen1999}: no configuration has both
lower SOP count and higher macro-F1 (Fig.~\ref{fig:pareto27}). These are
operation-count comparisons, not measured hardware energy.

\subsection{Spike-timing mechanism}\label{sec:res-timing}

\emph{(RQ3, timing)} At the main $T=25$ operating point,
Table~\ref{tab:timing} shows a positive mapping contribution on all five
confirmation protocols, ranging from \FNDmapMin{} to \FNDmapMax{}~pp macro-F1.
The spreading contribution ranges from \FNDspreadMin{} to \FNDspreadMax{}~pp
and carries every cross-protocol sign change in the total timing premium. 
The \FNTimingSpread{}-pp range of $\dtime$ corresponds to
\FNDmapRange{}~pp\footnote{Computed from stored full-precision values;
recomputing the mapping range from the two-decimal cells of
Table~\ref{tab:timing} can differ by 0.01~pp.} of mapping variation and
\FNDspreadRange{}~pp of spreading variation. On NSL-KDD
and KDDCup99, the mapping and spreading terms have similar magnitude and
opposite signs, leaving a total timing premium near zero.

\begin{table}[!ht]
\centering
\caption{Three-arm timing decomposition with \texttt{LeakyParallel} at
% CR13: Explain the mapped-arm mean in its caption
$T=25$, \rev{seeds 42--44. The mapped arm reuses this subset of the benchmark
predictions; Table~\ref{tab:strict-results} averages all five seeds.} Differences
are in pp of macro-F1. Control diagnostics are reported in Table~\ref{tab:a2checks}.}
\label{tab:timing}
\tabtiny
\setlength{\tabcolsep}{2.0pt}
\begin{tabular}{@{}lrrrrrr@{}}
\toprule
\textbf{Protocol} & \textbf{Map+spread} & \textbf{Spread} & \textbf{No spread} & \textbf{$\dmap$} & \textbf{$\dspread$} & \textbf{$\dtime$} \\
\midrule
% CR39: Use the supplied frozen macros for all timing-arm means
CTU-13 f0   & \rev{$\FNArmLatCtuA$} & \rev{$\FNArmPermCtuA$} & \rev{$\FNArmDeltaCtuA$} & \FNDmapCtuA & \FNDspreadCtuA & \FNTimingCtuA \\
CIC-IDS2017 & \rev{$\FNArmLatCic$} & \FNArmPermCic & \rev{$\FNArmDeltaCic$} & \FNDmapCic & \FNDspreadCic & \FNTimingCic \\
NSL-KDD     & \rev{$\FNArmLatNsl$} & \FNArmPermNsl & \rev{$\FNArmDeltaNsl$} & \FNDmapNsl & \FNDspreadNsl & \FNTimingNsl \\
KDDCup99    & \rev{$\FNArmLatKdd$} & \rev{$\FNArmPermKdd$} & \rev{$\FNArmDeltaKdd$} & \FNDmapKdd & \FNDspreadKdd & \FNTimingKdd \\
CTU-13 f1   & \rev{$\FNArmLatCtuB$} & \rev{$\FNArmPermCtuB$} & \rev{$\FNArmDeltaCtuB$} & \FNDmapCtuB & \FNDspreadCtuB & \FNTimingCtuB \\
\bottomrule
\end{tabular}
% CR57: State the assumption needed for the component-bound interpretation.
\tabnote{\rev{If residual mapping information benefits the spread-only arm,
the KDDCup99 mapping estimate is a lower bound and the spreading estimate is
the corresponding upper bound.}}
\end{table}

The manipulation checks support this decomposition. The active-feature sets and
per-timestep spike-count profiles match between the mapped and spread-only
controls. In the mapped arm, the feature-magnitude--time association is
approximately Spearman $\rho=\FNSpearmanLatency$ on four protocols; KDDCup99
is the exception at $\rho=\FNSpearmanLatencyLo$ because discretization at
$T=25$ produces more tied spike times.\footnote{The mapped-arm correlations
across the five protocols are $-0.990$, $-0.860$, $-0.997$, $-0.996$, and
$-0.998$; KDDCup99 is the $-0.860$ case.} Permutation substantially reduces 
 this association, although KDDCup99 retains
residual correlation of approximately $-0.210$
(Table~\ref{tab:perm-diagnostic}). \rev{The bound interpretation of its
\FNDmapKdd{}~pp mapping and \FNDspreadKdd{}~pp spreading estimates therefore
requires the residual information to benefit the spread-only arm.}

% CR14: Qualify the auxiliary directional hypothesis
This decomposition is post hoc and contributes to no model ranking.
\rev{The auxiliary directional hypothesis was formulated after related results
were available. On CTU-13 fold~0 at $T=25$, the mapping component accounts for
\FNATwoPredPass\% of the total timing premium, with the remainder attributable
to temporal spreading. We report these components descriptively rather than
as an independent confirmatory test of encoding superiority.}

% CR58: Identify Pearson correlations and avoid inferring stability from a pooled r.
\rev{Across $T\in\{5,10,25,50\}$, the pooled Pearson correlation between $T$
and $\dtime$ is $r=\FNTsweepR$ across \FNTsweepCells{} protocol--$T$ cells.
This weak pooled association does not establish within-protocol stability:
CTU-13 fold~1 changes timing-effect sign across $T$ (Table~\ref{tab:tsweep}).
Under that table's range-to-SE criterion, $\dmap$ varies on \FNMapMovesN{}
protocols and $\dspread$ on \FNSpreadMovesN{}. The four-point Pearson
correlation between the components is \FNRhoKdd{} for KDDCup99,
\FNRhoCic{} for CIC-IDS2017, \FNRhoCtuA{} for CTU-13 fold~0, and
\FNRhoCtuB{} for fold~1, with NSL-KDD the exception at $r=\FNRhoNsl$.
These negative correlations indicate compensating component changes, although
the total effect also varies within protocols. With only four paired values
per protocol, the correlations are descriptive.}

\paragraph{Resolution-dependent mapping}
CTU-13 fold~1 provides the clearest example. Its mapping contribution decreases
from $+50.46$~pp at $T=5$ to $+3.71$~pp at $T=25$ and $-0.87$~pp at $T=50$;
the final estimate is small relative to its three-seed uncertainty and is not
interpreted as evidence of a negative mapping contribution. At $T=5$, the
identical-permutation fraction is \FNCtuBIdenticalFrac{} and the resulting
amplitude--time correlation is \FNCtuBSpearmanAtwo{}, yet the spread-only arm
reaches only \FNCtuBAtwoTfive{} macro-F1, compared with
\FNCtuBAtwoTtwentyfive{} at $T=25$. On this protocol, the amplitude-to-time
association is therefore most consequential at coarse temporal resolution.

KDDCup99 has a different control limitation. The identical-permutation fraction
is \FNKddIdenticalFrac{}, and the spread-only control retains an
amplitude--time correlation of approximately $-0.210$.
\rev{If that residual information improves the spread-only arm's performance,
it biases $\dmap$ downward and $\dspread$ upward by equal amounts while
leaving $\dtime$ unchanged. The one-sided bounds apply under this assumption;
the residual correlation alone does not establish the performance effect.}

\paragraph{Distribution shift and timing sign}
Scalar train--test shift does not explain the sign of the timing premium.
CTU-13 fold~1 has the largest reported shift (median per-feature KS
\FNShiftKsCtuB{}, domain-classifier AUC \FNShiftAucCtuB{}), whereas the
KDD-family AUCs are \FNShiftAucKddLo{}--\FNShiftAucKddHi{}
(Table~\ref{tab:shift}). Across the five confirmation protocols, the rank
correlation between shift magnitude and $\dtime$ is \FNShiftRho{}. Moreover,
the two most-shifted CTU-13 folds lie at opposite ends of the timing result,
with \FNTimingCtuA{}~pp on fold~0 and \FNTimingCtuB{}~pp on fold~1. We
therefore do not attribute the timing sign to scalar distribution-shift
magnitude.

% =====================================================================
\section{Limitations and Threats to Validity}\label{sec:threats}

\paragraph{Scope and model comparison}
Our conclusions concern SNN design choices under a common architecture and
% CR26: Use primary-analysis terminology
training budget, not general SNN superiority. The \rev{two-axis analysis}
is the primary confirmation; the complete 27-configuration sweep was added
later, while the timing and distribution-shift analyses are post hoc. The
neuron-family ordering also reflects a shared operating point rather than
family-specific optimization. \texttt{SLSTM} and \texttt{SConv2dLSTM} require
a lower firing threshold because the library default produces no spikes, and
some CTU-13 fold~1 runs show optimization instability. In addition,
\texttt{LeakyParallel} is a parallelized variant of \texttt{Leaky}, so their
similar predictive performance should not be interpreted as evidence for
distinct neuron dynamics.

\paragraph{Dataset, transfer, and timing boundaries}
The confirmation protocols reduce identifiable train--test dependencies but
retain limitations of the underlying datasets. CIC-IDS2017 uses row order as a
proxy for temporal order and has no U2R support in the held-out interval; the
KDD-family benchmarks contain duplicates, rare classes, and attacks absent from
training; and CTU-13 measures scenario rather than unseen-host transfer because
hosts can recur across scenarios. The timing decomposition is interpretive
rather than causal and varies with temporal resolution. In particular, the
positive mapping contribution observed on all five protocols applies at
$T=25$, while KDDCup99 retains residual amplitude--time information after
% CR15: State the remaining generalization limits
permutation, \rev{so the component-bound interpretation depends on that
residual information benefiting the control}.
\rev{The full configuration ranking is evaluated at $T=25$; the
\texttt{LeakyParallel} timing sweep does not establish ranking stability across
timestep budgets. The findings are limited to the four datasets and static
flow representations studied. Generalization to native-label CIC-IDS2017 tasks,
other traffic distributions, architectures, or native event streams remains
untested.}

\paragraph{Computational activity}
SOPs are fanout-weighted operation counts rather than measurements of energy,
latency, or memory traffic. They are most comparable when the neuron model is
fixed; cross-family comparisons are less complete because cell-specific state
updates and recurrent projections are not fully represented. We therefore
interpret SOPs as computational-activity estimates rather than
hardware-efficiency measurements.

% =====================================================================
\section{Conclusion}\label{sec:conclusion}
We evaluated 27 SNN neuron--encoding configurations under screening and
% CR16: Align the conclusion with the qualified ranking
leakage-resistant confirmation. \rev{At the shared operating point,
\texttt{LeakyParallel/latency} achieved the lowest mean rank in both complete
evaluations but was not statistically separated from \texttt{Leaky/latency};
their screening order reversed when CTU-13 was excluded. The two rankings}
preserved most pairwise configuration orderings (Kendall $\tau=\FNTau$), although measured performance
changed under the stricter protocol, particularly on CTU-13. This supports
using screening for design selection and confirmation for performance
assessment.

% CR47: Condense the concluding synthesis
\rev{At $T=25$, amplitude-to-time mapping contributes \FNDmapMin{} to
\FNDmapMax{}~pp of macro-F1, while temporal spreading ranges from
\FNDspreadMin{} to \FNDspreadMax{}~pp and accounts for cross-protocol sign
changes in the total timing effect. Sparse input coding does not imply sparse
internal activity, motivating separate assessment of representation and
fanout-weighted SOPs. Future work should identify the traffic properties
governing these components and test their consistency across architectures
and hardware.}

\begin{comment}
    \section*{AI Disclosure}
    We used OpenAI's ChatGPT to assist with language editing and improving the clarity and organization of the manuscript text. The tool did not materially affect the study design, experimental data, analyses, or conclusions. The authors verified the correctness and originality of all content and take full responsibility for the manuscript.
\end{comment}

\section*{Acknowledgement}
The authors acknowledge research support from the Hewson Faculty Award (\# 1003-31880-214241) at the University of Alabama and NVIDIA Academic Grant Program. 

% CR63: Use the bundled bibliography with reference [2] corrected to article 774.
% This file preserves the 24 citation keys and displayed reference records.
\bibliographystyle{IEEEtran}
\bibliography{references}

% =====================================================================
% APPENDICES
% =====================================================================

\clearpage
\onecolumn
\appendices

% Single-column appendix float behavior
\renewcommand{\topfraction}{0.95}
\renewcommand{\bottomfraction}{0.95}
\renewcommand{\textfraction}{0.05}
\renewcommand{\floatpagefraction}{0.85}

\setcounter{topnumber}{5}
\setcounter{bottomnumber}{5}
\setcounter{totalnumber}{8}

\setlength{\textfloatsep}{10pt plus 2pt minus 2pt}
\setlength{\floatsep}{8pt plus 2pt minus 2pt}
\setlength{\intextsep}{8pt plus 2pt minus 2pt}

% =====================================================================
\section{Protocol, Duplication, and Metric Audits}
\label{app:audits}
% =====================================================================

In this appendix, we report the protocol, duplication, and metric audits that
support our leakage-resistant confirmation analysis.

\subsection{Split and Preprocessing Audit}

Table~\ref{tab:proto-audit} summarizes the evaluation artifacts identified
during screening and the corresponding controls applied during confirmation.
Each row states the issue, its measured extent, and its treatment under the
confirmation protocol.

\begin{table}[!htbp]
\centering
\caption{Protocol audit motivating the confirmation protocol.}
\label{tab:proto-audit}
\tabtiny
\begin{tabularx}{0.92\textwidth}{@{}lXXX@{}}
\toprule
\textbf{Dataset} & \textbf{Issue audited} & \textbf{Finding} &
\textbf{Treatment} \\
\midrule

NSL-KDD &
Exact train/test duplicates &
\FNDupNsl\% of test rows &
Duplicate-free sensitivity reported \\

KDDCup99 &
Exact train/test duplicates &
\FNDupKdd\% of test rows &
Duplicate-free sensitivity; ranking unchanged \\

CIC-IDS2017 &
Random split interleaves one attack burst &
Present in screening &
% CR71: The reported CIC-IDS2017 confirmation is within-capture order-disjoint.
\rev{Within-capture order-disjoint confirmation} \\

CTU-13 &
Whole-scenario host aggregates with random flow split &
$\FNCtuLegacy\rightarrow\FNCtuStrict$ macro-F1 &
Causal aggregates plus scenario-disjoint folds \\

All &
Categorical vocabulary fitted on train and test &
0 NSL-KDD / 2 KDDCup99 test rows affected &
Learned transforms fitted on training rows only \\

\bottomrule
\end{tabularx}
\end{table}

% =====================================================================
\subsection{Duplicate Sensitivity}
% =====================================================================

Table~\ref{tab:dedup} tests whether exact train--test duplication changes the
encoding-axis conclusions. Table~\ref{tab:labelmap} retains the fixed
five-class mapping used for the KDD-family protocols.

\par\medskip
\noindent
\begin{minipage}[t]{0.58\textwidth}
\vspace{0pt}
\centering

\captionof{table}{Attack-type to five-class mapping for NSL-KDD and KDDCup99,
applied identically under screening and confirmation.}
\label{tab:labelmap}

\tabtiny
\setlength{\tabcolsep}{3pt}
\renewcommand{\arraystretch}{1.03}

\begin{tabularx}{\linewidth}{@{}lX@{}}
\toprule
\textbf{Class} & \textbf{Attack labels} \\
\midrule

Normal &
\texttt{normal} \\

DoS &
\texttt{back}, \texttt{land}, \texttt{neptune}, \texttt{pod},
\texttt{smurf}, \texttt{teardrop}, \texttt{apache2},
\texttt{udpstorm}, \texttt{processtable}, \texttt{worm},
\texttt{mailbomb} \\

Probe &
\texttt{ipsweep}, \texttt{nmap}, \texttt{portsweep}, \texttt{satan},
\texttt{mscan}, \texttt{saint} \\

R2L &
\texttt{ftp\_write}, \texttt{guess\_passwd}, \texttt{imap},
\texttt{multihop}, \texttt{phf}, \texttt{spy},
\texttt{warezclient}, \texttt{warezmaster}, \texttt{sendmail},
\texttt{named}, \texttt{snmpgetattack}, \texttt{snmpguess},
\texttt{xlock}, \texttt{xsnoop}, \texttt{httptunnel} \\

U2R &
\texttt{buffer\_overflow}, \texttt{loadmodule}, \texttt{perl},
\texttt{rootkit}, \texttt{ps}, \texttt{sqlattack}, \texttt{xterm} \\

\bottomrule
\end{tabularx}

\tabnote{Assignments for \texttt{httptunnel}, \texttt{snmpgetattack},
\texttt{snmpguess}, \texttt{ps}, \texttt{sqlattack}, and \texttt{xterm}
differ across the literature. The mapping shown here reproduces the class
supports in Table~\ref{tab:perclass}; the R2L fallback was not invoked.}

\end{minipage}
\hfill
\begin{minipage}[t]{0.39\textwidth}
\vspace{0pt}
\centering

\captionof{table}{Duplicate-free sensitivity, encoding axis,
\FNSeeds{} seeds.}
\label{tab:dedup}

\tabtiny
\setlength{\tabcolsep}{2.5pt}
\renewcommand{\arraystretch}{1.03}

\begin{tabular}{@{}llrrrr@{}}
\toprule
\textbf{Protocol} & \textbf{Enc.} &
\textbf{Excl.} & \textbf{Official} &
\textbf{Dedup.} & $\Delta$ \\
& & \textbf{(\%)} & & & \textbf{(pp)} \\
\midrule

\multirow{3}{*}{NSL-KDD}
 & latency & \multirow{3}{*}{2.684} & 0.5846 & 0.5810 & $-0.364$ \\
 & rate    & & 0.5471 & 0.5431 & $-0.399$ \\
 & delta   & & 0.5197 & 0.5164 & $-0.335$ \\

\midrule
\multirow{3}{*}{KDDCup99}
 & latency & \multirow{3}{*}{\FNDupKdd} & 0.6309 & 0.6234 & $-0.753$ \\
 & rate    & & 0.6089 & 0.6009 & $-0.795$ \\
 & delta   & & 0.6083 & 0.5967 & $-1.159$ \\

\midrule
\multirow{3}{*}{CIC-IDS2017}
 & latency & \multirow{3}{*}{10.862} & 0.7842 & 0.7838 & $-0.039$ \\
 & rate    & & 0.7277 & 0.7452 & $+1.753$ \\
 & delta   & & 0.4981 & 0.4956 & $-0.250$ \\

\midrule
\multirow{3}{*}{CTU-13 f0}
 & latency & \multirow{3}{*}{0.001} & 0.9985 & 0.9985 & $+0.000$ \\
 & rate    & & 0.9876 & 0.9876 & $+0.000$ \\
 & delta   & & 0.5938 & 0.5938 & $+0.000$ \\

\midrule
\multirow{3}{*}{CTU-13 f1}
 & latency & \multirow{3}{*}{0.001} & 0.7702 & 0.7702 & $+0.000$ \\
 & rate    & & 0.7750 & 0.7751 & $+0.000$ \\
 & delta   & & 0.8825 & 0.8825 & $+0.001$ \\

\bottomrule
\end{tabular}

\tabnote{Removing duplicates changes absolute scores but preserves the
three-encoding ordering on every audited protocol. CIC-IDS2017 F1 was
recomputed on the fixed label universe; no ordering changes.}

\end{minipage}
\par\medskip

% CR17: Publish the CIC-IDS2017 mapping
\rev{Table~\ref{tab:cic-labelmap} documents the CIC-IDS2017 grouping used in
all protocols. The KDD-style names identify the implemented comparison task;
they do not establish equivalence to the KDD attack categories.}

\begin{table}[!htbp]
\centering
\caption{\rev{Study-specific CIC-IDS2017 label mapping, applied identically
% CR71: Same scope as Section III-B and the protocol-audit table.
under screening and within-capture order-disjoint confirmation.}}
\label{tab:cic-labelmap}
\tabtiny
\begin{tabularx}{0.92\textwidth}{@{}lX@{}}
\toprule
\rev{\textbf{Group}} & \rev{\textbf{Source labels}} \\
\midrule
\rev{Normal} & \rev{BENIGN} \\
\rev{DoS} & \rev{DDoS; DoS variants (Hulk, GoldenEye, slowloris, Slowhttptest); Bot} \\
\rev{Probe} & \rev{PortScan} \\
\rev{R2L} & \rev{FTP-Patator; SSH-Patator; Heartbleed} \\
\rev{U2R} & \rev{Web Attack variants (Brute Force, XSS, SQL Injection); Infiltration} \\
\bottomrule
\end{tabularx}
\tabnote{\rev{The loader strips surrounding whitespace and normalizes the
corrupted separator in Web Attack labels. Unrecognized labels fall back to
R2L. Records are not excluded by label conversion; duplicates are retained,
and infinite or missing feature values are replaced with zero.}}
\end{table}

\subsection{Metric Audit and Reproducibility}
\label{app:audit}

The order-disjoint CIC-IDS2017 test interval contains no U2R samples.
Accordingly, the primary metric remains macro-F1 over the fixed five-class task
universe, with the unsupported class contributing zero. A post-run audit found
that the initial implementation omitted the explicit label set, making the
averaging denominator prediction-dependent. We recomputed the affected metrics
from persisted predictions without retraining. One absolute value changed,
while every rank, mean rank, and hypothesis-test conclusion remained unchanged.

We record the amendment log, class-support audit, split-manifest hashes, and
source artifact for each reported value. The build regenerates the committed
macro file and requires byte-for-byte agreement before proceeding.

\FloatBarrier

% =====================================================================
\section{Design-Space Rankings}\label{app:screening}
% ======================================================================
% CR27: Use primary-analysis terminology
This appendix reports the screening population, \rev{primary} two-axis confirmation, and complete confirmation ranking summarized in Sections~\ref{sec:res-rank} and~\ref{sec:res-strict}.

\subsection{Screening Population}

Figure~\ref{fig:screening-rank} shows the 27-configuration ranking, while
Tables~\ref{tab:scr-rank} and~\ref{tab:scr-focused} summarize configurations
and paired comparisons. Table~\ref{tab:scr-noctu} repeats the ranking without
CTU-13 to test dependence on the dataset most affected by the protocol audit.

\par\medskip
\noindent
\begin{minipage}[t]{0.47\textwidth}
\vspace{0pt}
\centering

\includegraphics[
    width=\linewidth,
    height=0.50\textheight,
    keepaspectratio
]{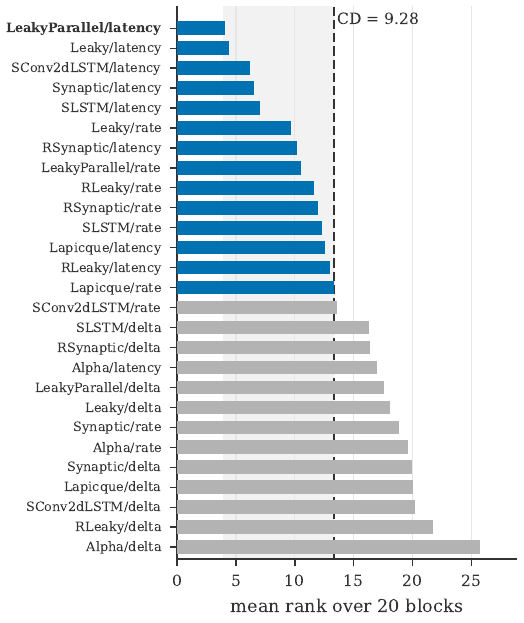}

\captionof{figure}{Screening design-space ranking over
\FNScrBlocks{} blocks with Nemenyi CD~\FNScrCD. Fourteen configurations lie
within one CD of the leader.}
\label{fig:screening-rank}

\end{minipage}
\hfill
\begin{minipage}[t]{0.51\textwidth}
\vspace{0pt}
\centering

\captionof{table}{Top 10 of the \FNScrVariants{} neuron-encoding variants by
mean rank over the \FNScrBlocks{} screening blocks.}
\label{tab:scr-rank}

\fontsize{6.4}{8.0}\selectfont
\setlength{\tabcolsep}{0.8pt}
\renewcommand{\arraystretch}{2.0}

\begin{tabular}{@{}lrcccc@{}}
\toprule
\textbf{Variant} & \textbf{Rank} &
\textbf{NSL-KDD} & \textbf{KDDCup99} &
\textbf{CIC-IDS2017} & \textbf{CTU-13} \\
\midrule
LeakyParallel/latency & 4.03$^{\dagger}$ &
$.5930\pm.0195$ & $.6199\pm.0065$ & $.9862\pm.0028$ & $1.0000\pm.0000$ \\

Leaky/latency & 4.40$^{\dagger}$ &
$.5742\pm.0176$ & $.6337\pm.0191$ & $.9850\pm.0021$ & $1.0000\pm.0000$ \\

SConv2dLSTM/latency & 6.15$^{\dagger}$ &
$.5693\pm.0113$ & $.6252\pm.0164$ & $.9808\pm.0034$ & $1.0000\pm.0000$ \\

Synaptic/latency & 6.53$^{\dagger}$ &
$.5804\pm.0178$ & $.6087\pm.0250$ & $.9794\pm.0063$ & $1.0000\pm.0000$ \\

SLSTM/latency & 7.00$^{\dagger}$ &
$.5772\pm.0217$ & $.6103\pm.0138$ & $.9824\pm.0026$ & $1.0000\pm.0000$ \\

Leaky/rate & 9.65$^{\dagger}$ &
$.5763\pm.0276$ & $.6189\pm.0172$ & $.8636\pm.0350$ & $0.9970\pm.0007$ \\

RSynaptic/latency & 10.18$^{\dagger}$ &
$.5240\pm.0267$ & $.5978\pm.0357$ & $.9831\pm.0048$ & $0.9999\pm.0000$ \\

LeakyParallel/rate & 10.55$^{\dagger}$ &
$.5812\pm.0238$ & $.6161\pm.0188$ & $.8530\pm.0354$ & $0.9971\pm.0007$ \\

RLeaky/rate & 11.65$^{\dagger}$ &
$.5494\pm.0378$ & $.6099\pm.0203$ & $.8585\pm.0354$ & $0.9972\pm.0007$ \\

RSynaptic/rate & 11.95$^{\dagger}$ &
$.5444\pm.0315$ & $.6093\pm.0283$ & $.8573\pm.0293$ & $0.9974\pm.0005$ \\
\bottomrule
\end{tabular}

\tabnote{$^{\dagger}$Not separated from the leader by the Nemenyi procedure
at $\alpha=0.05$; CD $=\FNScrCD$. The CTU-13 screening partition is
optimistic, as discussed in Table~\ref{tab:proto-audit}.}

\end{minipage}
\par\medskip

\par\medskip
\noindent
\begin{minipage}[t]{0.49\textwidth}
\vspace{0pt}
\centering

\captionof{table}{Focused paired screening comparisons over
\FNScrBlocks{} blocks.}
\label{tab:scr-focused}

\tabtiny
\renewcommand{\arraystretch}{2.0}
\begin{tabular}{@{}llcrrl@{}}
\toprule
\textbf{A} & \textbf{B} & \textbf{W/L/T} &
\textbf{Med.\ $\Delta$} & \textbf{Holm $p$} &
\textbf{Bootstrap $\Delta$} \\
\midrule

LP/latency & LP/rate
& \FNScrWinsA/\FNScrWinsB/0
& \FNScrMedDelta
& \FNScrHolmP
& $[\FNScrBootLo,\FNScrBootHi]$ \\

LP/latency & Leaky/latency
& 14/6/0
& +0.00
& \FNScrLeakyP$^{*}$
& $[-0.97,+1.38]$ \\

\bottomrule
\end{tabular}

\tabnote{LP denotes \texttt{LeakyParallel}.
$^{*}$Not significant at $\alpha=0.05$ after Holm.}

\end{minipage}
\hfill
\begin{minipage}[t]{0.49\textwidth}
\vspace{0pt}
\centering

\captionof{table}{Screening ranks over the
\FNNoCtuBlocks{} blocks excluding CTU-13.}
\label{tab:scr-noctu}

\tabtiny
\renewcommand{\arraystretch}{1.25}
\begin{tabular}{@{}r@{\hspace{16pt}}lr@{}}
\toprule
\textbf{\#} & \textbf{Variant} & \textbf{Mean rank} \\
\midrule
1  & \FNNoCtuLeader        & \FNNoCtuLeadRank \\
2  & LeakyParallel/latency & \FNNoCtuChampRank \\
3  & SConv2dLSTM/latency   & 6.40 \\
4  & Synaptic/latency      & 7.73 \\
5  & SLSTM/latency         & 8.20 \\
6  & Leaky/rate            & 8.80 \\
7  & LeakyParallel/rate    & 10.07 \\
8  & SLSTM/rate            & 10.73 \\
9  & RSynaptic/latency     & 11.40 \\
10 & RLeaky/rate           & 11.73 \\
\bottomrule
\end{tabular}

\end{minipage}
\par\medskip

% CR28: Use primary-analysis terminology
\subsection{\rev{Primary} Two-Axis Confirmation}

\begin{figure}[!htbp]
\centering
\includegraphics[width=\textwidth]{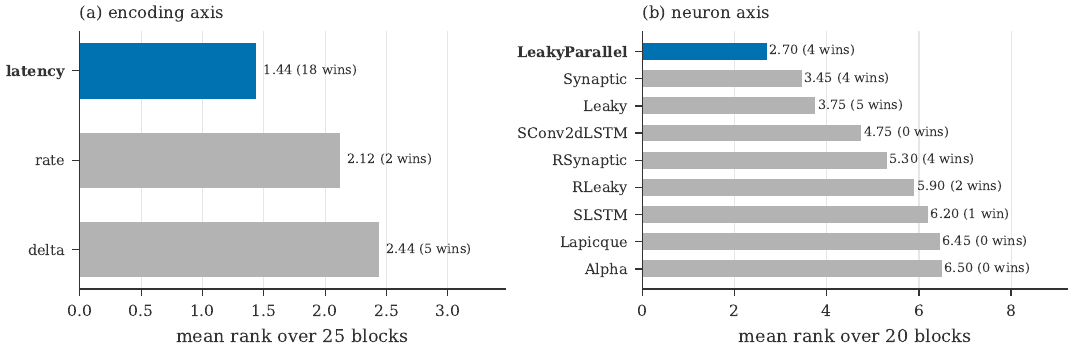}
% CR30: Use primary-analysis terminology
\caption{Leakage-resistant confirmation rankings for the \rev{primary} encoding
and neuron axes.}
\label{fig:confirmation-axes}
\end{figure}

% CR29: Use primary-analysis terminology
The two \rev{primary} axes hold one design factor fixed at a time.
Figure~\ref{fig:confirmation-axes} reports their mean ranks, and
Table~\ref{tab:neuron-results} gives the corresponding neuron-axis metrics.
The neuron ordering is interpreted only at the shared operating point of
Section~\ref{sec:enc}.

\begin{table}[!htbp]
\centering
\caption{Neuron-axis confirmation with latency encoding fixed, \FNSeeds{}
seeds (mean $\pm$ across-seed SD).}
\label{tab:neuron-results}
\tabsmall
\renewcommand{\arraystretch}{1.14}
\begin{tabular}{@{}l@{\hspace{8pt}}l@{\hspace{16pt}}c@{\hspace{16pt}}c@{\hspace{16pt}}c@{\hspace{16pt}}c@{\hspace{16pt}}r@{}}
\toprule
\textbf{Protocol} & \textbf{Neuron} & \textbf{Macro-F1} & \textbf{MCC} &
\textbf{DR} & \textbf{FAR} & \textbf{SOPs/sample} \\
\midrule

\multirow{9}{*}{NSL-KDD}
 & \textbf{LeakyParallel} & $\mathbf{0.5846 \pm 0.0279}$ & $0.6585 \pm 0.0272$ & $0.6696 \pm 0.0299$ & $0.0648 \pm 0.0145$ & 3{,}363 \\
 & Synaptic     & $0.5678 \pm 0.0198$ & $0.6517 \pm 0.0201$ & $0.6528 \pm 0.0207$ & $0.0626 \pm 0.0164$ & 2{,}505 \\
 & SConv2dLSTM  & $0.5639 \pm 0.0192$ & $0.6406 \pm 0.0161$ & $0.6267 \pm 0.0203$ & $0.0518 \pm 0.0175$ & 8{,}620 \\
 & Leaky        & $0.5637 \pm 0.0232$ & $0.6482 \pm 0.0205$ & $0.6354 \pm 0.0164$ & $0.0494 \pm 0.0103$ & 2{,}782 \\
 & SLSTM        & $0.5533 \pm 0.0155$ & $0.6436 \pm 0.0184$ & $0.6181 \pm 0.0221$ & $0.0372 \pm 0.0013$ & 9{,}003 \\
 & RSynaptic    & $0.5281 \pm 0.0306$ & $0.6397 \pm 0.0216$ & $0.6442 \pm 0.0226$ & $0.0514 \pm 0.0148$ & 4{,}377 \\
 & RLeaky       & $0.4829 \pm 0.0265$ & $0.6048 \pm 0.0194$ & $0.5940 \pm 0.0311$ & $0.0519 \pm 0.0233$ & 6{,}725 \\
 & Lapicque     & $0.4813 \pm 0.0018$ & $0.6280 \pm 0.0051$ & $0.6029 \pm 0.0051$ & $0.0490 \pm 0.0094$ & 2{,}968 \\
 & Alpha        & $0.4644 \pm 0.0056$ & $0.6028 \pm 0.0125$ & $0.5922 \pm 0.0073$ & $0.0652 \pm 0.0096$ & 2{,}409 \\

\midrule
\multirow{9}{*}{KDDCup99}
 & \textbf{Leaky} & $\mathbf{0.6575 \pm 0.0201}$ & $0.8266 \pm 0.0023$ & $0.9113 \pm 0.0012$ & $0.0138 \pm 0.0035$ & 2{,}406 \\
 & Synaptic      & $0.6322 \pm 0.0123$ & $0.8237 \pm 0.0027$ & $0.9108 \pm 0.0013$ & $0.0163 \pm 0.0006$ & 2{,}000 \\
 & SLSTM         & $0.6314 \pm 0.0120$ & $0.8228 \pm 0.0021$ & $0.9106 \pm 0.0018$ & $0.0145 \pm 0.0006$ & 7{,}261 \\
 & LeakyParallel & $0.6309 \pm 0.0119$ & $0.8269 \pm 0.0032$ & $0.9122 \pm 0.0010$ & $0.0130 \pm 0.0049$ & 2{,}843 \\
 & SConv2dLSTM   & $0.6072 \pm 0.0284$ & $0.8242 \pm 0.0046$ & $0.9098 \pm 0.0019$ & $0.0061 \pm 0.0032$ & 7{,}227 \\
 & Lapicque      & $0.5985 \pm 0.0096$ & $0.8252 \pm 0.0032$ & $0.9096 \pm 0.0007$ & $0.0116 \pm 0.0044$ & 2{,}699 \\
 & RSynaptic     & $0.5871 \pm 0.0374$ & $0.8167 \pm 0.0037$ & $0.9081 \pm 0.0026$ & $0.0113 \pm 0.0030$ & 5{,}004 \\
 & RLeaky        & $0.5597 \pm 0.0438$ & $0.8206 \pm 0.0026$ & $0.9099 \pm 0.0031$ & $0.0075 \pm 0.0045$ & 5{,}863 \\
 & Alpha         & $0.5434 \pm 0.0086$ & $0.8225 \pm 0.0026$ & $0.9094 \pm 0.0013$ & $0.0147 \pm 0.0010$ & 2{,}227 \\

\midrule
\multirow{9}{*}{CTU-13 f0}
 & \textbf{Synaptic} & $\mathbf{0.9995 \pm 0.0001}$ & $0.9990 \pm 0.0002$ & $0.9995 \pm 0.0001$ & $0.0005 \pm 0.0003$ & 3{,}408 \\
 & LeakyParallel & $0.9985 \pm 0.0013$ & $0.9971 \pm 0.0026$ & $0.9981 \pm 0.0024$ & $0.0008 \pm 0.0009$ & 3{,}399 \\
 & Leaky         & $0.9963 \pm 0.0043$ & $0.9927 \pm 0.0086$ & $0.9941 \pm 0.0072$ & $0.0002 \pm 0.0001$ & 3{,}117 \\
 & Alpha         & $0.9937 \pm 0.0032$ & $0.9874 \pm 0.0064$ & $0.9899 \pm 0.0055$ & $0.0006 \pm 0.0003$ & 3{,}056 \\
 & RLeaky        & $0.9821 \pm 0.0128$ & $0.9650 \pm 0.0246$ & $0.9719 \pm 0.0219$ & $0.0026 \pm 0.0028$ & 4{,}076 \\
 & Lapicque      & $0.9815 \pm 0.0110$ & $0.9638 \pm 0.0212$ & $0.9692 \pm 0.0185$ & $0.0004 \pm 0.0002$ & 3{,}244 \\
 & SConv2dLSTM   & $0.9775 \pm 0.0429$ & $0.9585 \pm 0.0779$ & $0.9620 \pm 0.0732$ & $0.0005 \pm 0.0004$ & 7{,}939 \\
 & RSynaptic     & $0.9665 \pm 0.0123$ & $0.9353 \pm 0.0232$ & $0.9440 \pm 0.0211$ & $0.0007 \pm 0.0006$ & 4{,}284 \\
 & SLSTM         & $0.9028 \pm 0.0325$ & $0.8214 \pm 0.0514$ & $0.8419 \pm 0.0597$ & $0.0103 \pm 0.0060$ & 12{,}309 \\

\midrule
\multirow{9}{*}{CTU-13 f1}
 & \textbf{RSynaptic} & $\mathbf{0.9114 \pm 0.1090}$ & $0.8285 \pm 0.2158$ & $0.9026 \pm 0.1125$ & $0.0763 \pm 0.1353$ & 4{,}114 \\
 & RLeaky        & $0.8984 \pm 0.0827$ & $0.8151 \pm 0.1337$ & $0.9360 \pm 0.0780$ & $0.1372 \pm 0.1786$ & 4{,}259 \\
 & LeakyParallel & $0.7702 \pm 0.0041$ & $0.6220 \pm 0.0046$ & $0.9968 \pm 0.0025$ & $0.4562 \pm 0.0088$ & 3{,}084 \\
 & Alpha         & $0.7608 \pm 0.0037$ & $0.6033 \pm 0.0116$ & $0.9903 \pm 0.0076$ & $0.4669 \pm 0.0028$ & 2{,}885 \\
 & Synaptic      & $0.6502 \pm 0.2004$ & $0.3581 \pm 0.4108$ & $0.7852 \pm 0.3670$ & $0.4675 \pm 0.0016$ & 3{,}119 \\
 & Lapicque      & $0.6307 \pm 0.1252$ & $0.2885 \pm 0.2672$ & $0.7272 \pm 0.2443$ & $0.4613 \pm 0.0031$ & 2{,}990 \\
 & SConv2dLSTM   & $0.5799 \pm 0.2792$ & $0.1895 \pm 0.5524$ & $0.5801 \pm 0.4324$ & $0.3921 \pm 0.0855$ & 7{,}493 \\
 & Leaky         & $0.5288 \pm 0.1538$ & $0.0717 \pm 0.2939$ & $0.5372 \pm 0.2714$ & $0.4602 \pm 0.0023$ & 2{,}878 \\
 & SLSTM         & $0.3161 \pm 0.1373$ & $-0.3332 \pm 0.2554$ & $0.1809 \pm 0.2283$ & $0.4703 \pm 0.0026$ & 11{,}336 \\

\bottomrule
\end{tabular}

\tabnote{The neuron axis uses one shared operating point. Several CTU-13
fold~1 cells show optimization failure or large across-seed variability and
are not interpreted as evidence of intrinsic neuron-family superiority.}
\end{table}

\FloatBarrier

\subsection{Complete Confirmation Ranking}

Table~\ref{tab:conf27} places the complete screening and confirmation rankings
side by side, while Fig.~\ref{fig:rank-shift} visualizes their displacement.

% CR59: Supply the exact finite-depth definition behind the frozen RBO scores.
\rev{For the complete lists, the reported RBO is
\[
  \mathrm{RBO}_{\FNConfVariants}(p)
  =(1-p)\sum_{d=1}^{\FNConfVariants}p^{d-1}A_d,
\]
where $A_d$ is the fraction of configurations shared by the two top-$d$ sets.
The sum is neither normalized by its finite-depth weight nor supplemented
with a tail-extrapolation term.}

\par\medskip
\noindent
\begin{minipage}[t]{0.47\textwidth}
\vspace{0pt}
\centering

\captionof{table}{Complete confirmation ranking over
\FNConfBlocks{} paired protocol--seed blocks beside the screening ranking.}
\label{tab:conf27}

\fontsize{8.0}{8.0}\selectfont
\setlength{\tabcolsep}{2.0pt}
\renewcommand{\arraystretch}{1.40}

\begin{tabular}{@{}rlrrrr@{}}
\toprule
\textbf{Pos.} & \textbf{Variant} & \textbf{Conf.} &
\textbf{Scr.} & \textbf{Scr.\ pos.} & \textbf{Move} \\
\midrule
1  & LeakyParallel/latency & 5.92  & 4.03  & 1  & 0 \\
2  & Leaky/latency         & 7.92  & 4.40  & 2  & 0 \\
3  & Synaptic/latency      & 8.32  & 6.53  & 4  & +1 \\
4  & SConv2dLSTM/latency   & 10.60 & 6.15  & 3  & -1 \\
5  & Leaky/rate            & 10.64 & 9.65  & 6  & +1 \\
6  & LeakyParallel/rate    & 11.40 & 10.55 & 8  & +2 \\
7  & RSynaptic/rate        & 11.48 & 11.95 & 10 & +3 \\
8  & RSynaptic/latency     & 12.16 & 10.18 & 7  & -1 \\
9  & SLSTM/rate            & 12.68 & 12.30 & 11 & +2 \\
10 & SLSTM/latency         & 12.72 & 7.00  & 5  & -5 \\
11 & SConv2dLSTM/rate      & 13.04 & 13.55 & 15 & +4 \\
12 & RLeaky/latency        & 13.24 & 12.95 & 13 & +1 \\
13 & RLeaky/rate           & 13.72 & 11.65 & 9  & -4 \\
14 & SLSTM/delta           & 14.20 & 16.25 & 16 & +2 \\
15 & Lapicque/rate         & 14.44 & 13.30 & 14 & -1 \\
16 & RSynaptic/delta       & 14.52 & 16.38 & 17 & +1 \\
17 & Leaky/delta           & 14.92 & 18.10 & 20 & +3 \\
18 & SConv2dLSTM/delta     & 15.32 & 20.20 & 25 & +7 \\
19 & Lapicque/latency      & 15.48 & 12.53 & 12 & -7 \\
20 & LeakyParallel/delta   & 15.92 & 17.60 & 19 & -1 \\
21 & Alpha/latency         & 16.64 & 16.95 & 18 & -3 \\
22 & Synaptic/delta        & 16.72 & 19.98 & 23 & +1 \\
23 & Synaptic/rate         & 17.60 & 18.80 & 21 & -2 \\
24 & Lapicque/delta        & 18.36 & 20.00 & 24 & 0 \\
25 & Alpha/rate            & 18.76 & 19.60 & 22 & -3 \\
26 & RLeaky/delta          & 20.20 & 21.70 & 26 & 0 \\
27 & Alpha/delta           & 21.08 & 25.75 & 27 & 0 \\
\bottomrule
\end{tabular}

% CR33: Correct math delimiters in top-k overlaps
\tabnote{Top-$k$ overlap is \rev{$\FNTopKthree/3$,
$\FNTopKfive/5$, $\FNTopKten/10$, and $\FNTopKfourteen/14$}.
Restricted Kendall $\tau=\FNTauTop$ for the leading
\FNTauTopN{} configurations and $\FNTauBottom$ for the remaining
\FNTauBottomN{}.}

\end{minipage}
\hfill
\begin{minipage}[t]{0.50\textwidth}
\vspace{0pt}
\centering

\includegraphics[
  width=\linewidth,
  height=0.70\textheight,
  keepaspectratio
]{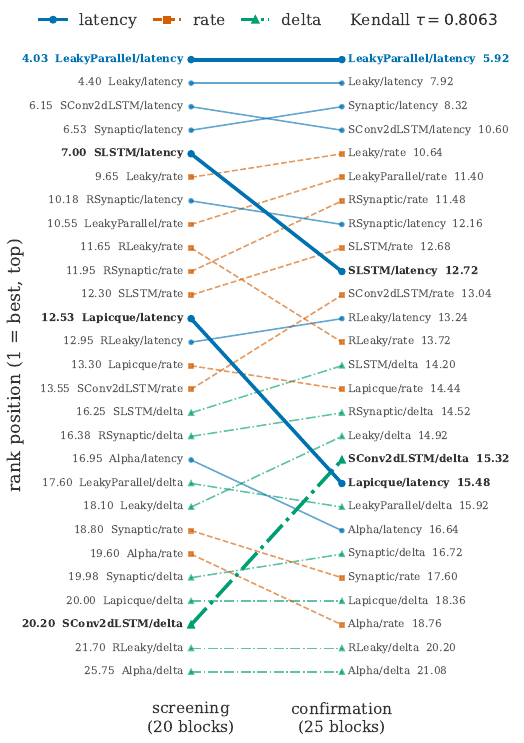}

\captionof{figure}{Rank displacement between screening and confirmation for
all \FNScrVariants{} configurations. Kendall $\tau=\FNTau$.}
\label{fig:rank-shift}

\end{minipage}
\par\medskip

\FloatBarrier

% =====================================================================
\section{Encoder and Neural-Code Measurements}\label{app:encoder}
% =====================================================================
This appendix reports the encoder characterization, temporal-readout,
timing-decomposition, and activity measurements supporting
Sections~\ref{sec:res-activity} and~\ref{sec:res-timing}.

\subsection{Encoder Characterization}

Tables~\ref{tab:sparsity} and~\ref{tab:input-spikes} characterize the sparsity
and input-spike behavior of the three encoders before network dynamics are
considered.

\par\medskip
\noindent
\begin{minipage}[t]{0.56\textwidth}
\vspace{0pt}
\centering

% CR34: Define the pre-gate active-feature count
\captionof{table}{Feature sparsity and encoder gate mass under confirmation.
\rev{Counts are mean features per sample; ``Active'' denotes $x>0$ before gating.}}
\label{tab:sparsity}

\tabtiny
\renewcommand{\arraystretch}{1.25}
\begin{tabular}{@{}lrrrrrrr@{}}
\toprule
\textbf{Protocol} & $d$ & \textbf{Zero} & \textbf{Active}
& $(0,\theta_\mathrm{lat}]$
& $(\theta_\mathrm{lat},\theta_\delta]$
& \textbf{Band (\%)} & $>\theta_\delta$ \\
\midrule

NSL-KDD
& \FNDimNsl & \FNZeroNsl & \FNNslLatInSpk
& \FNBandNsl & \FNGateBandNsl & \FNGateBandPctNsl & 13.560 \\

KDDCup99
& \FNDimKdd & \FNZeroKdd & 12.491
& \FNBandKdd & \FNGateBandKdd & \FNGateBandPctKdd & 12.086 \\

CIC-IDS2017
& \FNDimCic & \FNZeroCic & 39.968
& \FNBandCic & \FNGateBandCic & \FNGateBandPctCic & 37.961 \\

CTU-13 f0
& \FNDimCtuA & \FNZeroCtuA & 22.859
& \FNBandCtuA & \FNGateBandCtuA & \FNGateBandPctCtuA & 21.362 \\

CTU-13 f1
& \FNDimCtuB & \FNZeroCtuB & 20.999
& \FNBandCtuB & \FNGateBandCtuB & \FNGateBandPctCtuB & 19.467 \\

\bottomrule
\end{tabular}

\end{minipage}
\hfill
\begin{minipage}[t]{0.40\textwidth}
\vspace{0pt}
\centering

\captionof{table}{Input spikes per sample by protocol and encoding,
\texttt{LeakyParallel}.}
\label{tab:input-spikes}

\tabtiny
\renewcommand{\arraystretch}{1.25}
\begin{tabular}{@{}lrrrr@{}}
\toprule
\textbf{Protocol} & \textbf{delta} & \textbf{latency} &
\textbf{rate} & \textbf{rate/lat.} \\
\midrule
NSL-KDD     & 13.560 & 14.580 & 238.585 & 16.36 \\
KDDCup99    & 12.086 & 12.491 & 245.577 & 19.66 \\
CIC-IDS2017 & 37.961 & 39.904 & 369.202 & 9.25 \\
CTU-13 f0   & 21.362 & 22.233 & 309.056 & 13.90 \\
CTU-13 f1   & 19.467 & 20.652 & 308.258 & 14.93 \\
\bottomrule
\end{tabular}

\tabnote{Latency and delta are deterministic given $x$; rate redraws per seed.}

\end{minipage}
\par\medskip

\subsection{Gate Controls and Temporal Readout}
Table~\ref{tab:gatecontrol} tests whether the latency--delta comparison is
sensitive to their different amplitude gates, while Table~\ref{tab:readout}
measures how quickly latency and rate approach their respective terminal
macro-F1 values. Because $T_q$ is defined relative to each configuration's own
terminal score, it measures convergence rather than terminal performance.
The two thresholds also need not agree: on CTU-13 fold~0, rate reaches
$T_{95}$ first, whereas latency reaches $T_{99}$ first. We therefore report
both thresholds.

\par\medskip
\noindent
\begin{minipage}[t]{0.55\textwidth}
\vspace{0pt}
\centering

% CR18: Specify the gate-control seed IDs
\captionof{table}{Encoder gate controls,
\texttt{LeakyParallel}, \FNEoneSeeds{} seeds \rev{(42--44)}.}
\label{tab:gatecontrol}

\tabtiny
\renewcommand{\arraystretch}{1.25}
\begin{tabular}{@{}lrrrrrr@{}}
\toprule
& \multicolumn{2}{c}{\textbf{$\theta_\mathrm{lat}=0$}}
& \multicolumn{2}{c}{\textbf{Baseline}}
& \multicolumn{2}{c}{\textbf{$\theta_\mathrm{lat}=\FNDeltaThresh$}} \\
\cmidrule(lr){2-3}\cmidrule(lr){4-5}\cmidrule(lr){6-7}
\textbf{Protocol}
& \textbf{F1} & \textbf{in spk}
& \textbf{F1} & \textbf{in spk}
& \textbf{F1} & \textbf{in spk} \\
\midrule
NSL-KDD     & 0.5818 & 14.5802 & 0.5818 & 14.5802 & 0.5878 & 13.5603 \\
KDDCup99    & 0.6394 & 12.4906 & 0.6394 & 12.4906 & 0.6456 & 12.0861 \\
CIC-IDS2017 & 0.7846 & 39.9681 & 0.7825 & 39.9044 & 0.7848 & 37.9611 \\
CTU-13 f0   & 0.9990 & 22.8591 & 0.9995 & 22.2328 & 0.9986 & 21.3625 \\
CTU-13 f1   & 0.7678 & 20.9995 & 0.7710 & 20.6520 & 0.7717 & 19.4672 \\
\bottomrule
\end{tabular}
% CR35: Define gate-control baseline and retained-feature counts
\tabnote{\rev{``Baseline'' uses $\theta_\mathrm{lat}=\FNLatThresh$.
``in spk'' is the mean number of retained features, and thus input spikes,
per sample.}}

% CR65: Suppress interword glue between the adjacent appendix tables.
\end{minipage}%
\hfill
\begin{minipage}[t]{0.45\textwidth}
\vspace{0pt}
\centering

\captionof{table}{Median $T_{95}$ and $T_{99}$ by confirmation protocol.}
\label{tab:readout}

\tabtiny
\renewcommand{\arraystretch}{1.25}
\begin{tabular}{@{}llrrrrl@{}}
\toprule
& & \multicolumn{2}{c}{$T_{95}$}
& \multicolumn{2}{c}{$T_{99}$} & \\
\cmidrule(lr){3-4}\cmidrule(lr){5-6}
\textbf{Protocol} & \textbf{Best}
& lat. & rate & lat. & rate & \textbf{Earlier} \\
\midrule

NSL-KDD & latency
& \FNTNFNslLat & \FNTNFNslRat
& \textbf{\FNTNNNslLat} & \FNTNNNslRat & latency \\

KDDCup99 & latency
& \FNTNFKddLat & \FNTNFKddRat
& \FNTNNKddLat & \textbf{\FNTNNKddRat} & rate \\

CIC-IDS2017 & latency
& \FNTNFCicLat & \FNTNFCicRat
& \FNTNNCicLat & \textbf{\FNTNNCicRat} & rate \\

CTU-13 f0 & latency
& \FNTNFCtuALat & \FNTNFCtuARat
& \textbf{\FNTNNCtuALat} & \FNTNNCtuARat & latency \\

CTU-13 f1 & delta
& \FNTNFCtuBLat & \FNTNFCtuBRat
& \FNTNNCtuBLat & \textbf{\FNTNNCtuBRat} & rate \\

\bottomrule
\end{tabular}

\end{minipage}
\par\medskip

\subsection{Timing-Decomposition Controls}
Tables~\ref{tab:a2checks}--\ref{tab:tsweep} report the controls behind the
timing decomposition. The matched-input arms preserve the active-feature set,
while the permutation diagnostic measures residual amplitude--time
association. Together, these checks verify that the decomposition changes the
temporal organization of a fixed spike set rather than the features permitted
to spike. Table~\ref{tab:tsweep} then reports the two components and their sum
across the four timestep settings.

\par\medskip
\noindent
\begin{minipage}[t]{0.5\textwidth}
\vspace{0pt}
\centering

\captionof{table}{Matched-input checks for the timing decomposition.}
\label{tab:a2checks}

\tabtiny
\renewcommand{\arraystretch}{1.25}
\begin{tabular}{@{}lrrr@{}}
\toprule
\textbf{Check} & \textbf{Mapped+spread} &
\textbf{Spread only} & \textbf{No spread} \\
\midrule
Differing active-feature cells & 0 & 0 & 0 \\
Active-feature multiset & identical & identical & identical \\
Spike profile vs.\ mapped & reference & identical & concentrated \\
Feature value--time Spearman $\rho$
& $\approx\FNSpearmanLatency^\dagger$
& $\FNSpearmanSpread$ & n/a \\
\bottomrule
\end{tabular}

% CR60: Identify the aggregation of the displayed correlation summaries.
\tabnote{\rev{Correlations summarize cross-protocol medians.}
$^\dagger$Approximately $\FNSpearmanLatency$ on four protocols;
KDDCup99 is the exception at $\FNSpearmanLatencyLo$.
\rev{Its spread-only residual is $\FNSpearmanSpreadLo$.}}

\end{minipage}
\hfill
\begin{minipage}[t]{0.48\textwidth}
\vspace{0pt}
\centering

\captionof{table}{Spread-only permutation diagnostics.}
\label{tab:perm-diagnostic}

\tabtiny
\renewcommand{\arraystretch}{1.25}
% CR61: Omit the unreconciled feature/bin diagnostic, retaining the other checks.
% AUTHOR CHECK: The omitted Feat./bin values were 4.43, 0.89, 2.92, 0.46
% (first value: FNCtuBTieDensity). Full-test retained features / T instead
% gives approximately 4.13, 0.83, 2.92, 0.50. Restore the column only after
% confirming its diagnostic population and denominator from the saved output.
\begin{tabular}{@{}lrrr@{}}
\toprule
\textbf{Protocol} & $T$ & \textbf{Same time} & $\rho_{\mathrm{amp,time}}$ \\
\midrule
CTU-13 f1 & 5  & 0.307 & \FNCtuBSpearmanAtwo \\
CTU-13 f1 & 25 & 0.089 & $-0.027$ \\
NSL-KDD   & 5  & 0.438 & $-0.049$ \\
KDDCup99  & 25 & \FNKddIdenticalFrac & $-0.210$ \\
\bottomrule
\end{tabular}
\tabnote{\rev{Correlations are Spearman coefficients.}}

\end{minipage}
\par\medskip

\begin{table}[!htbp]
\centering
% CR19: Specify timestep-sweep checkpoint provenance
\caption{Timing components across the timestep budget,
\texttt{LeakyParallel}, \FNEoneSeeds{} seeds \rev{(42--44)}. Values are macro-F1 differences
% CR62: The stored component correlations are Pearson, not Spearman.
in pp; \rev{$r$ is the Pearson correlation between $\dmap$ and $\dspread$}
across the four $T$ settings.
\rev{Mapped models are trained separately at $T=5,10,50$; $T=25$ reuses the
benchmark predictions.}}
\label{tab:tsweep}

\tabtiny
\renewcommand{\arraystretch}{1.25}

\begin{tabular*}{0.90\textwidth}{
@{\extracolsep{\fill}}
l
l
@{\hspace{5pt}}
r r r r
@{\hspace{7pt}}
r
@{\hspace{7pt}}
l
@{\hspace{7pt}}
r
@{}
}
\toprule
\textbf{Protocol} & &
$T{=}5$ & $10$ & $25$ & $50$ &
\textbf{Mean} & \textbf{Verdict} & \rev{$r$} \\
\midrule

& $\dmap$
& $+1.30$ & $+5.89$ & $\FNDmapNsl$ & $+6.92$
& $\FNMapMeanNsl$ & stable & \\

NSL-KDD
& $\dspread$
& $-3.76$ & $-3.98$ & $\FNDspreadNsl$ & $-0.73$
& $\FNSpreadMeanNsl$ & stable
& \raisebox{1.4ex}[0pt][0pt]{$\FNRhoNsl$} \\

& $\dtime$
& $-2.46$ & $+1.92$ & $\FNDtimeNsl$ & $+6.19$
& $\FNDtimeMeanNsl$ & -- & \\

\midrule

& $\dmap$
& $+0.67$ & $+5.17$ & $\FNDmapKdd$ & $+2.12$
& $\FNMapMeanKdd$ & varies & \\

KDDCup99
& $\dspread$
& $-2.01$ & $-4.02$ & $\FNDspreadKdd$ & $-2.37$
& $\FNSpreadMeanKdd$ & varies
& \raisebox{1.4ex}[0pt][0pt]{$\FNRhoKdd$} \\

& $\dtime$
& $-1.33$ & $+1.15$ & $\FNDtimeKdd$ & $-0.25$
& $\FNDtimeMeanKdd$ & -- & \\

\midrule

& $\dmap$
& $+23.96$ & $+13.87$ & $\FNDmapCic$ & $+16.30$
& $\FNMapMeanCic$ & varies & \\

CIC-IDS2017
& $\dspread$
& $+10.69$ & $+14.57$ & $\FNDspreadCic$ & $+12.20$
& $\FNSpreadMeanCic$ & stable
& \raisebox{1.4ex}[0pt][0pt]{$\FNRhoCic$} \\

& $\dtime$
& $+34.65$ & $+28.44$ & $\FNDtimeCic$ & $+28.50$
& $\FNDtimeMeanCic$ & -- & \\

\midrule

& $\dmap$
& $+31.59$ & $+25.30$ & $\FNDmapCtuA$ & $+21.29$
& $\FNMapMeanCtuA$ & varies & \\

CTU-13 f0
& $\dspread$
& $+14.95$ & $+34.00$ & $\FNDspreadCtuA$ & $+35.11$
& $\FNSpreadMeanCtuA$ & stable
& \raisebox{1.4ex}[0pt][0pt]{$\FNRhoCtuA$} \\

& $\dtime$
& $+46.54$ & $+59.30$ & $\FNDtimeCtuA$ & $+56.40$
& $\FNDtimeMeanCtuA$ & -- & \\

\midrule

& $\dmap$
& $+50.46$ & $+38.36$ & $\FNDmapCtuB$ & $-0.87$
& $\FNMapMeanCtuB$ & varies & \\

CTU-13 f1
& $\dspread$
& $-43.67$ & $-53.83$ & $\FNDspreadCtuB$ & $+5.70$
& $\FNSpreadMeanCtuB$ & varies
& \raisebox{1.4ex}[0pt][0pt]{$\FNRhoCtuB$} \\

& $\dtime$
& $+6.79$ & $-15.47$ & $\FNDtimeCtuB$ & $+4.83$
& $\FNDtimeMeanCtuB$ & -- & \\

\bottomrule
\end{tabular*}

% CR36: Distinguish exact timing identities from rounded display values
% CR69: Match the inclusive threshold implemented in the repo.
\tabnote{\rev{``Varies'' means that the across-$T$ range is at least twice
the largest per-cell SE.} \rev{The identity $\dtime=\dmap+\dspread$ holds before rounding;
independent rounding can produce 0.01-pp differences in the displayed cells.}
\rev{KDDCup99's component estimates are one-sided bounds if residual mapping
information benefits the spread-only arm.}}
\end{table}

\subsection{Activity and Operation Decomposition}
\begin{figure}[!htbp]
\centering

\begin{minipage}[t]{0.48\textwidth}
\centering
\includegraphics[width=\linewidth]{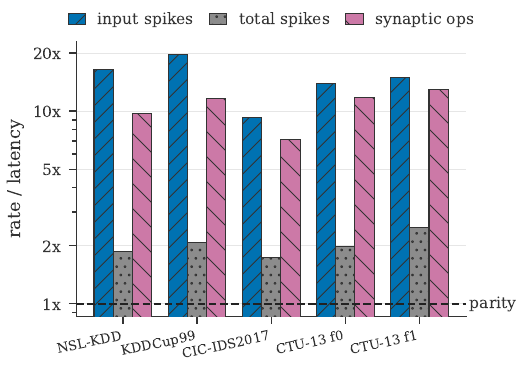}
\end{minipage}
\hfill
\begin{minipage}[t]{0.48\textwidth}
\centering
\includegraphics[width=\linewidth]{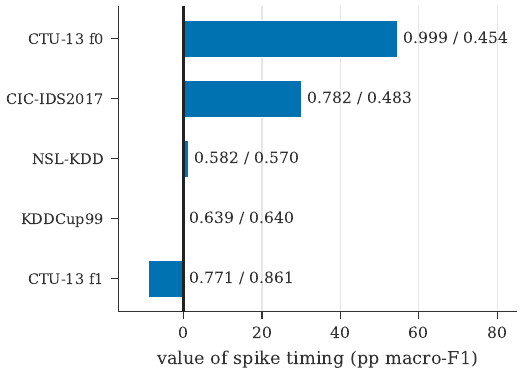}
\end{minipage}

\caption{Supporting neural-code measurements. Left: rate-to-latency activity
ratios. Right: total timing premium $\dtime$ at matched input activity.}
\label{fig:code-support}
\end{figure}

Table~\ref{tab:sopdecomp} separates raw spike activity from the operations
driven by those spikes. Figure~\ref{fig:code-support} summarizes the
cross-protocol activity and timing measurements.

\begin{table}[!htbp]
\centering
% CR70: Replace the old single-seed attribution with the repo aggregation.
\caption{Per-layer spikes and SOPs for \texttt{LeakyParallel},
\rev{mean over five seeds}.}
\label{tab:sopdecomp}

\tabsmall
\renewcommand{\arraystretch}{1.50}
\begin{tabular}{@{}l@{\hspace{8pt}}l@{\hspace{8pt}}r@{\hspace{8pt}}r@{\hspace{8pt}}r@{\hspace{8pt}}r@{\hspace{8pt}}r@{\hspace{8pt}}r@{\hspace{8pt}}r@{\hspace{8pt}}c@{\hspace{8pt}}c@{}}
\toprule
& & \multicolumn{4}{c}{\textbf{Spikes per sample}} &
\multicolumn{3}{c}{\textbf{Synaptic operations}} &
\multicolumn{2}{c}{\textbf{Shares (\%)}} \\
\cmidrule(lr){3-6}\cmidrule(lr){7-9}\cmidrule(lr){10-11}

\textbf{Protocol} & \textbf{Enc.} &
\textbf{input} & \textbf{hidden} & \textbf{output} & \textbf{total} &
\textbf{input} & \textbf{hidden} & \textbf{total} &
\textbf{hidden spk} & \textbf{input SOP} \\
\midrule

\multirow{3}{*}{NSL-KDD}
 & delta   & 13.560  & 137.091 & 16.267 & 166.918 & 1{,}736  & 685   & 2{,}421  & 82.1 & 71.7 \\
 & latency & 14.580  & 299.341 & 32.885 & 346.806 & 1{,}866  & 1{,}497 & 3{,}363  & 86.3 & 55.5 \\
 & rate    & 238.585 & 392.635 & 16.791 & 648.010 & 30{,}539 & 1{,}963 & 32{,}502 & 60.6 & 94.0 \\

\midrule
\multirow{3}{*}{KDDCup99}
 & delta   & 12.086  & 74.535  & 16.138 & 102.758 & 1{,}547  & 373   & 1{,}920  & 72.5 & 80.6 \\
 & latency & 12.491  & 248.876 & 37.698 & 299.065 & 1{,}599  & 1{,}244 & 2{,}843  & 83.2 & 56.2 \\
 & rate    & 245.577 & 358.167 & 18.467 & 622.211 & 31{,}434 & 1{,}791 & 33{,}225 & 57.6 & 94.6 \\

\midrule
\multirow{3}{*}{CIC-IDS2017}
 & delta   & 37.961  & 80.791  & 15.713 & 134.465 & 4{,}859  & 404   & 5{,}263  & 60.1 & 92.3 \\
 & latency & 39.904  & 341.944 & 24.436 & 406.284 & 5{,}108  & 1{,}710 & 6{,}817  & 84.2 & 74.9 \\
 & rate    & 369.202 & 315.230 & 19.617 & 704.049 & 47{,}258 & 1{,}576 & 48{,}834 & 44.8 & 96.8 \\

\midrule
\multirow{3}{*}{CTU-13 f0}
 & delta   & 21.362  & 97.875  & 21.851 & 141.088 & 2{,}734  & 196   & 2{,}930  & 69.4 & 93.3 \\
 & latency & 22.233  & 276.445 & 25.927 & 324.605 & 2{,}846  & 553   & 3{,}399  & 85.2 & 83.7 \\
 & rate    & 309.056 & 318.020 & 16.987 & 644.064 & 39{,}559 & 636   & 40{,}195 & 49.4 & 98.4 \\

\midrule
\multirow{3}{*}{CTU-13 f1}
 & delta   & 19.467  & 156.979 & 28.569 & 205.016 & 2{,}492  & 314   & 2{,}806  & 76.6 & 88.8 \\
 & latency & 20.652  & 220.053 & 16.870 & 257.575 & 2{,}643  & 440   & 3{,}084  & 85.4 & 85.7 \\
 & rate    & 308.258 & 313.731 & 18.230 & 640.219 & 39{,}457 & 627   & 40{,}085 & 49.0 & 98.4 \\

\bottomrule
\end{tabular}

% CR37: State that the SOP tolerance applies before rounding
\tabnote{\rev{Input plus hidden SOPs reproduce the total to within
$\FNSopResidual$ before rounding; displayed integer components can differ from
their total by one operation.} The two share columns use different denominators.}
\end{table}

\FloatBarrier

% =====================================================================
\section{Boundary Conditions}
\label{app:boundary}
% =====================================================================

This appendix examines two conditions that affect how the reported
classification scores should be interpreted in deployment: transfer of a
validation-selected operating point and the class composition underlying
aggregate macro-F1 differences. These analyses do not alter the configuration
rankings; they characterize how those rankings translate into particular
decision thresholds and classes.

\subsection{Operating-Point Transfer}

The primary results use the argmax decision rule because it provides a common
model-comparison criterion. A deployed detector, however, may instead select a
threshold to satisfy an operational false-alarm constraint. We therefore
calibrate each \texttt{LeakyParallel/latency} model on validation data to a
nominal $1\%$ FAR and freeze that threshold before test. This experiment asks
whether an operating point selected on validation retains the same meaning on
the held-out distribution.

\par\medskip
\noindent
\begin{minipage}[t]{0.43\textwidth}
\vspace{0pt}
\centering

\captionof{table}{Validation-to-test transfer of a $1\%$ FAR operating point
for \texttt{LeakyParallel/latency}.}
\label{tab:far}

\tabsmall
\renewcommand{\arraystretch}{1.50}
\begin{tabular}{@{}lrrr@{}}
\toprule
\textbf{Protocol} & \textbf{Val.\ FAR (\%)} &
\textbf{Test FAR (\%)} & \textbf{DR (\%)} \\
\midrule
CIC-IDS2017 & \FNFarValCic  & \FNFarTestCic  & \FNFarDrCic \\
KDDCup99    & \FNFarValKdd  & \FNFarTestKdd  & \FNFarDrKdd \\
NSL-KDD     & \FNFarValNsl  & \FNFarTestNsl  & \FNFarDrNsl \\
CTU-13 f0   & \FNFarValCtuA & \FNFarTestCtuA & \FNFarDrCtuA \\
CTU-13 f1   & \FNFarValCtuB & \FNFarTestCtuB & \FNFarDrCtuB \\
\bottomrule
\end{tabular}

\tabnote{Validation FAR remains within $0.4$~pp of the target on every
protocol.}

\end{minipage}
\hfill
\begin{minipage}[t]{0.53\textwidth}
\vspace{0pt}
\centering

\includegraphics[width=0.92\linewidth]{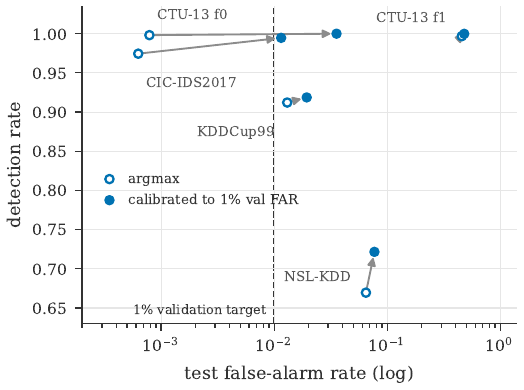}

\captionof{figure}{Shift from the argmax operating point
(Table~\ref{tab:strict-results}) to the calibrated point
(Table~\ref{tab:far}).}
\label{fig:fartransfer}

\end{minipage}
\par\smallskip

The validation results first confirm that the requested operating point can be
realized: all five validation FARs lie within $0.4$~pp of the $1\%$ target.
The test behavior is different. CIC-IDS2017 remains close to the calibrated
region at \FNFarTestCic\% FAR, and KDDCup99 rises to \FNFarTestKdd\%.
Larger departures appear on NSL-KDD and CTU-13 fold~0, while CTU-13 fold~1
reaches \FNFarTestCtuB\% FAR despite using a threshold selected at
\FNFarValCtuB\% validation FAR.

Because the threshold is frozen before test, these differences cannot be
attributed to test-set retuning. Instead, they measure whether the validation
score distribution transfers to the held-out data. Figure~\ref{fig:fartransfer}
also shows that moving from argmax to the calibrated threshold increases DR on
all five protocols, but the cost in FAR differs markedly. In particular, the
high DR on CTU-13 fold~1 should not be interpreted independently of its
corresponding FAR: the threshold detects nearly all attacks by also labeling a
large fraction of normal traffic as malicious.

The argmax and calibrated results therefore answer different questions.
Argmax supports a common comparison of predictive behavior across
configurations, whereas the calibrated experiment tests whether a fixed
validation-derived operating constraint transfers to test. The latter is
stable on some protocols and not on others. We do not interpret the $1\%$
target itself as an optimal deployment setting; it serves as a common reference
for exposing threshold-transfer sensitivity under distribution change.

\subsection{Per-Class Performance}

Table~\ref{tab:perclass} reports class-specific performance, while
Table~\ref{tab:margin-decomp} shows how individual classes contribute to the
latency-minus-rate macro-F1 difference. The KDD-family margins are concentrated
in R2L and U2R, whereas the CIC-IDS2017 difference is driven mainly by DoS and
R2L. The decomposition therefore identifies where an aggregate encoding
% CR31: Use primary-analysis terminology
difference originates without changing the \rev{fixed-class metric}.

% Centered compact table: the minipage occupies 58% of the page width.
\begin{center}
\begin{minipage}{0.5\textwidth}
\centering

\captionof{table}{Class contributions to the latency-minus-rate macro-F1
margin, \texttt{LeakyParallel}, \FNSeeds{} seeds. Entries are in pp.}
\label{tab:margin-decomp}

\tabsmall
\renewcommand{\arraystretch}{1.50}
\begin{tabular}{@{}lrrrrrr@{}}
\toprule
\textbf{Protocol} & $\Delta$\textbf{macro} &
\textbf{normal} & \textbf{DoS} & \textbf{Probe} &
\textbf{R2L} & \textbf{U2R} \\
\midrule

NSL-KDD &
$+3.75$ & $+0.02$ & $-0.03$ & $-0.04$ & $+2.24$ & $+1.56$ \\

KDDCup99 &
% CR66: Repo class contribution is 2.104 pp; round once from full precision.
$+2.21$ & $-0.03$ & $-0.00^{\dagger}$ & $-0.14$ & $+0.27$ & \rev{$+2.10$} \\

CIC-IDS2017 &
% CR67: Repo total is 5.654 pp; do not sum already-rounded class entries.
\rev{$+5.65$} & $+0.06$ & $+1.32$ & $+0.03$ & $+4.25$ & $0.00$ \\

\bottomrule
\end{tabular}

\tabnote{$^{\dagger}$KDDCup99 DoS contributes
$-3.7\times10^{-4}$~pp and is treated as tied.
\rev{Components and totals are rounded independently from full precision.}}

\end{minipage}
\end{center}

\begin{table}[!htbp]
\centering
% CR38: Identify the per-class aggregation and uncertainty
\caption{Per-class confirmation performance,
\texttt{LeakyParallel}, \FNSeeds{} seeds. Precision and recall are shown for
latency; F1 is shown for all three encodings.
\rev{F1 entries with uncertainty report mean $\pm$ across-seed SD;
precision and recall are seed means.}}
\label{tab:perclass}
\label{tab:perclass-enc}

\tabsmall
\renewcommand{\arraystretch}{1.0}
\begin{tabular}{@{}l@{\hspace{16pt}}l@{\hspace{16pt}}r@{\hspace{16pt}}r@{\hspace{16pt}}r@{\hspace{16pt}}r@{\hspace{16pt}}r@{\hspace{16pt}}r@{}}
\toprule
& & &
\multicolumn{2}{c}{\textbf{latency}} &
\multicolumn{3}{c}{\textbf{F1 by encoding}} \\
\cmidrule(lr){4-5}\cmidrule(lr){6-8}

\textbf{Protocol} & \textbf{Class} & \textbf{Support} &
\textbf{Prec.} & \textbf{Rec.} &
\textbf{latency} & \textbf{rate} & \textbf{delta} \\
\midrule

\multirow{6}{*}{NSL-KDD}
 & Normal & \FNSupNslNormal & 0.682 & 0.935
 & $\FNLatNslNormal\pm\FNLatNslNormalSd$
 & $\FNRateNslNormal\pm\FNRateNslNormalSd$
 & $\FNDeltaNslNormal\pm\FNDeltaNslNormalSd$ \\

 & DoS & \FNSupNslDos & 0.955 & 0.820
 & $\FNLatNslDos\pm\FNLatNslDosSd$
 & $\FNRateNslDos\pm\FNRateNslDosSd$
 & $\FNDeltaNslDos\pm\FNDeltaNslDosSd$ \\

 & Probe & \FNSupNslProbe & 0.715 & 0.669
 & $\FNLatNslProbe\pm\FNLatNslProbeSd$
 & $\FNRateNslProbe\pm\FNRateNslProbeSd$
 & $\FNDeltaNslProbe\pm\FNDeltaNslProbeSd$ \\

 & R2L & \FNSupNslRtl & 0.916 & \FNNslRtlRec
 & $\FNLatNslRtl\pm\FNLatNslRtlSd$
 & $\FNRateNslRtl\pm\FNRateNslRtlSd$
 & $\FNDeltaNslRtl\pm\FNDeltaNslRtlSd$ \\

 & U2R & \FNSupNslUtr & 0.668 & \FNNslUtrRec
 & $\FNLatNslUtr\pm\FNLatNslUtrSd$
 & $\FNRateNslUtr\pm\FNRateNslUtrSd$
 & $\FNDeltaNslUtr\pm\FNDeltaNslUtrSd$ \\

\cmidrule(l){2-8}
 & \textit{macro} & \FNNslTestN & & &
 $\mathit{\FNNslMacro}$ &
 $\mathit{\FNNslMacroRate}$ &
 $\mathit{\FNNslMacroDelta}$ \\

\midrule
\multirow{6}{*}{KDDCup99}
 & Normal & \FNSupKddNormal & 0.731 & 0.987
 & $\FNLatKddNormal\pm\FNLatKddNormalSd$
 & $\FNRateKddNormal\pm\FNRateKddNormalSd$
 & $\FNDeltaKddNormal\pm\FNDeltaKddNormalSd$ \\

 & DoS & \FNSupKddDos & 0.998 & 0.973
 & $\FNLatKddDos\pm\FNLatKddDosSd$
 & $\FNRateKddDos\pm\FNRateKddDosSd$
 & $\FNDeltaKddDos\pm\FNDeltaKddDosSd$ \\

 & Probe & \FNSupKddProbe & 0.787 & 0.768
 & $\FNLatKddProbe\pm\FNLatKddProbeSd$
 & $\FNRateKddProbe\pm\FNRateKddProbeSd$
 & $\FNDeltaKddProbe\pm\FNDeltaKddProbeSd$ \\

 & R2L & \FNSupKddRtl & 0.913 & 0.061
 & $\FNLatKddRtl\pm\FNLatKddRtlSd$
 & $\FNRateKddRtl\pm\FNRateKddRtlSd$
 & $\FNDeltaKddRtl\pm\FNDeltaKddRtlSd$ \\

 & U2R & \FNSupKddUtr & 0.558 & 0.383
 & $\FNLatKddUtr\pm\FNLatKddUtrSd$
 & $\FNRateKddUtr\pm\FNRateKddUtrSd$
 & $\FNDeltaKddUtr\pm\FNDeltaKddUtrSd$ \\

\cmidrule(l){2-8}
 & \textit{macro} & \FNKddTestN & & &
 $\mathit{\FNKddMacro}$ &
 $\mathit{\FNKddMacroRate}$ &
 $\mathit{\FNKddMacroDelta}$ \\

\midrule
\multirow{6}{*}{CIC-IDS2017}
 & Normal & \FNSupCicNormal & 0.998 & 0.999
 & $\FNLatCicNormal\pm\FNLatCicNormalSd$
 & $\FNRateCicNormal\pm\FNRateCicNormalSd$
 & $\FNDeltaCicNormal\pm\FNDeltaCicNormalSd$ \\

 & DoS & \FNSupCicDos & 0.980 & 0.950
 & $\FNLatCicDos\pm\FNLatCicDosSd$
 & $\FNRateCicDos\pm\FNRateCicDosSd$
 & $\FNDeltaCicDos\pm\FNDeltaCicDosSd$ \\

 & Probe & \FNSupCicProbe & 1.000 & 0.994
 & $\FNLatCicProbe\pm\FNLatCicProbeSd$
 & $\FNRateCicProbe\pm\FNRateCicProbeSd$
 & $\FNDeltaCicProbe\pm\FNDeltaCicProbeSd$ \\

 & R2L & \FNSupCicRtl & 0.988 & 0.935
 & $\FNLatCicRtl\pm\FNLatCicRtlSd$
 & $\FNRateCicRtl\pm\FNRateCicRtlSd$
 & $\FNDeltaCicRtl\pm\FNDeltaCicRtlSd$ \\

 & U2R & \FNSupCicUtr & -- & --
 & $\FNLatCicUtr$ & $\FNRateCicUtr$ & $\FNDeltaCicUtr$ \\

\cmidrule(l){2-8}
 & \textit{macro} & \FNCicTestN & & &
 $\mathit{\FNCicFixedLat}$ &
 $\mathit{\FNCicFixedRat}$ &
 $\mathit{\FNCicFixedDel}$ \\

\midrule
\multirow{3}{*}{CTU-13 f0}
 & Normal & \FNSupCtuANormal & 0.997 & 0.999
 & $\FNLatCtuANormal\pm\FNLatCtuANormalSd$
 & $\FNRateCtuANormal\pm\FNRateCtuANormalSd$
 & $\FNDeltaCtuANormal\pm\FNDeltaCtuANormalSd$ \\

 & Botnet & \FNSupCtuABotnet & 0.999 & 0.998
 & $\FNLatCtuABotnet\pm\FNLatCtuABotnetSd$
 & $\FNRateCtuABotnet\pm\FNRateCtuABotnetSd$
 & $\FNDeltaCtuABotnet\pm\FNDeltaCtuABotnetSd$ \\

\cmidrule(l){2-8}
 & \textit{macro} & \FNCtuATestN & & &
 $\mathit{\FNCtuAEncLat}$ &
 $\mathit{\FNCtuAEncRat}$ &
 $\mathit{\FNCtuAEncDel}$ \\

\midrule
\multirow{3}{*}{CTU-13 f1}
 & Normal & \FNSupCtuBNormal & 0.993 & 0.544
 & $\FNLatCtuBNormal\pm\FNLatCtuBNormalSd$
 & $\FNRateCtuBNormal\pm\FNRateCtuBNormalSd$
 & $\FNDeltaCtuBNormal\pm\FNDeltaCtuBNormalSd$ \\

 & Botnet & \FNSupCtuBBotnet & 0.722 & 0.997
 & $\FNLatCtuBBotnet\pm\FNLatCtuBBotnetSd$
 & $\FNRateCtuBBotnet\pm\FNRateCtuBBotnetSd$
 & $\FNDeltaCtuBBotnet\pm\FNDeltaCtuBBotnetSd$ \\

\cmidrule(l){2-8}
 & \textit{macro} & \FNCtuBTestN & & &
 $\mathit{\FNCtuBEncLat}$ &
 $\mathit{\FNCtuBEncRat}$ &
 $\mathit{\FNCtuBEncDel}$ \\

\bottomrule
\end{tabular}

\tabnote{F1 is the mean of per-seed F1 values and therefore need not equal the
harmonic mean of the displayed seed-mean precision and recall. CIC-IDS2017 U2R
has zero test support and contributes zero to all encodings.}
\end{table}

\FloatBarrier

% =====================================================================
\section{Distribution-Shift Diagnostic}
\label{app:shift}
% =====================================================================

This post hoc diagnostic tests whether train--test distribution shift explains
the sign of the timing premium. We report all five confirmation protocols and
the two additional CTU-13 folds.

\par\smallskip
\noindent
\begin{minipage}[t]{0.51\textwidth}
\vspace{0pt}
\centering

\captionof{table}{Train--test shift by protocol, ordered by
domain-classifier AUC.}
\label{tab:shift}

\tabtiny
\renewcommand{\arraystretch}{1.5}
\begin{tabular}{@{}lrrrrrrr@{}}
\toprule
\textbf{Protocol} & $d$ & \textbf{KS med.} & $W_{p90}$ &
\textbf{AUC} & \textbf{top-5 (\%)} &
$n_{W>0.1}$ & $\dtime$ \\
\midrule

CTU-13 f1
& \FNDimCtuB & \FNShiftKsCtuB & 0.2425 & \FNShiftAucCtuB
& 42.83 & 12 & \FNTimingCtuB \\

CTU-13 f0
& \FNDimCtuA & 0.3156 & 0.1590 & 0.9319
& 43.46 & 8 & \FNTimingCtuA \\

CTU-13 f3
& \FNDimCtuD & 0.2482 & 0.1165 & 0.9233
& 44.94 & 4 & -- \\

CTU-13 f2
& \FNDimCtuC & 0.1678 & 0.0833 & 0.9084
& 40.27 & 1 & -- \\

CIC-IDS2017
& \FNDimCic & 0.1848 & 0.1534 & 0.7572
& 17.15 & 26 & \FNTimingCic \\

NSL-KDD
& \FNDimNsl & 0.0030 & 0.0657 & 0.7499
& 34.82 & 9 & \FNTimingNsl \\

KDDCup99
& \FNDimKdd & 0.0002 & 0.0577 & 0.7184
& 35.96 & 5 & \FNTimingKdd \\

\bottomrule
\end{tabular}

\tabnote{KS is the median per-feature two-sample statistic; $W_{p90}$ is the
90th-percentile Wasserstein distance. Across the five confirmation protocols,
the rank correlation between shift magnitude and $\dtime$ is
$\FNShiftRho$.}

\end{minipage}
\hfill
\begin{minipage}[t]{0.45\textwidth}
\vspace{0pt}
\centering

\includegraphics[width=0.90\linewidth]{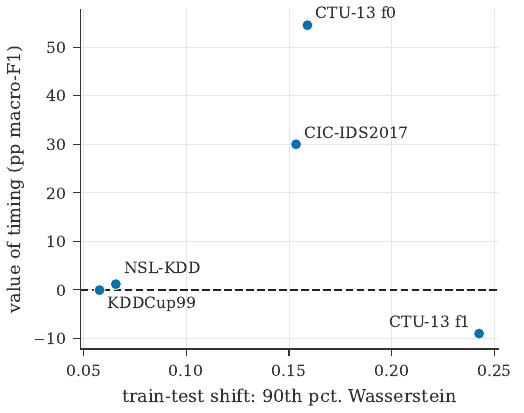}

\captionof{figure}{Distributional shift versus total timing premium. The two
most-shifted confirmation protocols lie at opposite extremes of $\dtime$
($\rho=\FNShiftRho$).}
\label{fig:shift}

\end{minipage}
\par\smallskip

Three observations follow. CTU-13 fold~1 has the largest measured shift, yet
fold~0 is also highly shifted and produces the opposite timing outcome.
Across the five confirmation protocols, shift magnitude and $\dtime$ have rank
correlation $\FNShiftRho$. Occupancy shifts can also exceed magnitude shifts,
suggesting that shift structure may matter beyond a single scalar distance.

\FloatBarrier

\end{document}